%% file: elsarticle/manuscript.tex
\documentclass[preprint,12pt]{elsarticle}
\usepackage[a4paper, margin=1in]{geometry}
\usepackage{url}
\usepackage{hyperref}
\usepackage{esvect}
\usepackage{float}
\usepackage{xcolor}
\usepackage{latexsym}
\usepackage{amssymb}
\usepackage{amsmath}
\usepackage{microtype} 

\usepackage{booktabs}
\usepackage{multirow}
\usepackage{siunitx}
\usepackage{placeins}
\usepackage{float}
\usepackage[table]{xcolor}

\begin{document}

\begin{frontmatter}

\title{Spatially Regularized Super-Resolved Constrained Spherical Deconvolution (SR$^2$-CSD) of Diffusion MRI Data}

\author[1]{Ekin Taskin\corref{cor1}}
\ead{ekin.taskin@epfl.ch}

\author[1]{Gabriel Girard}

\author[1]{Juan Luis Villarreal Haro}

\author[1,2]{Jonathan Rafael-Patiño}

\author[3]{Eleftherios Garyfallidis}

\author[1,2,4]{Jean-Philippe Thiran\fnref{fn1}}

\author[1,5,6]{Erick Jorge Canales-Rodríguez\fnref{fn1}\corref{cor1}}

\cortext[cor1]{Corresponding authors:}
\ead{erick.canalesrodriguez@epfl.ch}
\fntext[fn1]{Co-last senior authors}

\address[1]{Signal Processing Laboratory 5 (LTS5), École Polytechnique Fédérale de Lausanne (EPFL), Lausanne, Switzerland}
\address[2]{Radiology Department, Centre Hospitalier Universitaire Vaudois (CHUV) and University of Lausanne (UNIL), Lausanne, Switzerland}
\address[3]{Intelligent Systems Engineering, Indiana University, Bloomington, United States}
\address[4]{Center for Biomedical Imaging (CIBM), Lausanne, Switzerland}
\address[5]{Department of Signal Theory, Networking and Communications, University of Granada, Granada, Spain}
\address[6]{Andalusian Research Institute in Data Science and Computational Intelligence (DaSCI), University of Granada, Spain}

\begin{abstract}
Constrained Spherical Deconvolution (CSD) is widely used to estimate the white matter fiber orientation distribution (FOD) from diffusion MRI data. Its angular resolution depends on the maximum spherical harmonic order ($l_{max}$): low $l_{max}$ yields smooth but poorly resolved FODs, while high $l_{max}$, as in Super-CSD, enables resolving fiber crossings with small inter-fiber angles but increases sensitivity to noise. In this proof-of-concept study, we introduce Spatially Regularized Super-Resolved CSD (SR$^2$-CSD), a novel method that regularizes Super-CSD using a spatial FOD prior estimated via a self-calibrated total variation denoiser. We evaluated SR$^2$-CSD against CSD and Super-CSD across four datasets: (i) the HARDI-2013 challenge numerical phantom, assessing angular and peak number errors across multiple signal-to-noise ratio (SNR) levels and CSD variants (single-/multi-shell, single-/multi-tissue); (ii) the Sherbrooke in vivo dataset, evaluating spatial coherence of FODs; (iii) a six-subject test–retest dataset acquired with both full (96 gradient directions) and subsampled (45 directions) protocols, assessing reproducibility; and (iv) the DiSCo phantom, evaluating tractography accuracy under varying SNR levels and multiple noise repetitions. Across all evaluations, SR$^2$-CSD consistently reduced angular and peak number errors, improved spatial coherence, enhanced test–retest reproducibility, and yielded connectivity matrices more strongly correlated with ground-truth. Most improvements were statistically significant under multiple-comparison correction. These results demonstrate that incorporating spatial priors into CSD is feasible, mitigates estimation instability, and improves FOD reconstruction accuracy.
\end{abstract}


\end{frontmatter}

\section{Introduction}
\label{sec1}

Diffusion Magnetic Resonance Imaging (dMRI) is sensitive to the microscopic displacements of water molecules in biological tissues. This capability makes it a powerful tool for non-invasively probing the complex tissue architecture of the brain’s white matter (WM). Numerous techniques have been developed to estimate local WM fiber orientations \cite{alexander2005multiple}, including diffusion tensor imaging \cite{basser1994mr}, diffusion spectrum imaging \cite{wedeen2005mapping, Canales_DSI}, q-ball imaging \cite{tuch2004q, Descoteaux2007, Canales2009_eqbi, Aganj2010, Tristan2010}, diffusion orientation transform \cite{Evren2006, Canales2010_dot}, higher-order tensor models \cite{ozarslan2003generalized, liu2004characterizing, jensen2005diffusional}, and multi-tensor fitting algorithms \cite{tuch2002high}. Each of these approaches has specific strengths and limitations. Among them, spherical deconvolution (SD) methods \cite{tournier2004direct, Alexander2005, Alonso2007, Vemuri2007, tournier2007robust, acqua2007, Kaden2007, Descoteaux2009, Patel2010, acqua2010, tournier2011diffusion, Kaden2012, Sotiropoulos2012, daducci2013quantitative, Yeh2013, Parker2013, Canales2015_rumba, Canales2019_sparse, Lin2019, DeLuca2020, Elaldi2021} have emerged as the leading approach to estimate the voxel-wise fiber orientation distribution (FOD). These methods provide superior angular resolution for detecting fiber crossings at small inter-fiber angles, using single-shell \cite{tournier2007robust} or multi-shell \cite{Jeurissen2014} high angular resolution diffusion imaging (HARDI) data \cite{tuch2002high}.

Among the various parametric and non-parametric SD techniques, constrained spherical deconvolution (CSD) \cite{tournier2007robust}, as implemented in MRtrix3 \cite{MRtrix3} and Diffusion Imaging in Python (DIPY) \cite{garyfallidis2014dipy}, is the most widely adopted in both research and clinical practice. This method represents the FOD using orthogonal spherical harmonics (SH), providing a compact, efficient, and continuous representation of functions defined on the sphere. Importantly, CSD imposes a non-negativity constraint on the FOD to mitigate non-physical negative lobes. This constraint provides additional information that facilitates the implementation of super-resolution, producing a variant known as Super-CSD, in which the number of estimated SH coefficients exceeds the number of dMRI measurements \cite{tournier2007robust}.

Despite these advances, several limitations remain. A key parameter influencing the quality of the estimated FOD is the maximum SH order $l_{max}$ used to construct the deconvolution matrix, which determines the maximum achievable angular resolution. Low values (e.g., $l_{max}<8$) yield smoother and more stable estimates from noisy dMRI data by discarding higher frequency components. In contrast, higher $l_{max}$ values in Super-CSD enable sharper FODs with greater angular resolution, which is essential for resolving smaller inter-fiber angles. Although phantom experiments with Super-CSD and $l_{max}=16$ showed promising results \cite{Tournier2008_phantoms}, CSD with $l_{max}=8$ remains the recommended choice for real brain data \cite{tournier2007robust}. This indicates that the non-negativity constraint alone is insufficient for robust angular super-resolution. The deconvolution problem exhibits varying degrees of ill-conditioning, particularly for higher-order SH components or, equivalently, for higher values of the $l_{max}$ parameter. This ill-conditioning amplifies the impact of noise, so the higher angular resolution offered by Super-CSD often comes at the cost of stability, leading to spurious lobes and reduced consistency of orientations across adjacent voxels \cite{tournier2007robust}.

Several methods have been proposed in the literature that complement fiber orientation estimation with information about spatial correlations across voxels. These methods have shown greater stability than variants without such priors, yielding more spatially coherent and continuous voxel-to-voxel estimates with attenuated spurious fibers. Broadly, these approaches can be grouped into: (i) \textit{spatial regularization}, which enforces spatial smoothness \cite{Goh2009} and orientation continuity across neighboring voxels using penalties such as Total Variation (TV) \cite{Yuyuan2013, Canales2015_rumba} or fiber-continuity constraints \cite{Reisert2011, Reisert2013}; (ii) \textit{contextual flows and PDE-based enhancement}, typically applied as post-estimation refinement of estimated FODs (including CSD outputs) to improve alignment while preserving crossings \cite{Portegies2015, Nie2025}; (iii) \textit{neighborhood-aware, Bayesian schemes}, which couple voxelwise orientation estimation with spatial penalties and multi-resolution approaches \cite{Lenglet2018, Sotiropoulos2013, Sotiropoulos2016, Coronado2017, Auria2015}; and (iv) \textit{learning-based methods that leverage spatial context}, where convolutional neural networks or similar architectures use local 3D patches to improve spatial consistency of FOD \cite{Aguayo2024, Lucena2021, Lin2019, Hendriks2025, Elaldi2021}.

Inspired by these previous strategies that emphasize the importance of incorporating spatial information into diffusion modeling, we aim to extend the analytical structure of CSD by integrating spatial regularity directly within its formulation. Rather than redefining the classical method, this approach preserves its theoretical foundations while expanding its original formulation. Building on the strengths and widespread adoption of CSD, we introduce the Spatially Regularized Super-Resolved CSD (SR$^2$-CSD) method, which integrates spatial regularization as a prior alongside the non-negativity constraint. This formulation aims to enhance the stability and noise robustness of FOD estimation while preserving the interpretability and efficiency of conventional CSD. We therefore focus our evaluation on CSD and its Super-CSD variant to isolate the contribution of this spatial extension. We hypothesize that SR$^2$-CSD achieves the angular super-resolution of Super-CSD while providing greater spatial coherence, improved noise robustness, and higher reproducibility than both standard CSD and Super-CSD.

\section{Theory}
\label{sec2}

The single-shell single-tissue CSD (SSST-CSD) inverse problem \cite{tournier2007robust} is solved by using Tikhonov regularization, where the regularization term is designed to penalize negative FOD values without strictly forbidding them. An iterative refinement procedure is employed to improve the fitting sequentially. After obtaining an initial estimate of the FOD SH coefficients as a $p \times 1$ vector $\boldsymbol{f}$, the FOD amplitude $\boldsymbol{u}$ is evaluated along uniformly distributed unit vectors ${N}$ ($\sim$ 300). This is done using the $N \times p$ matrix $\boldsymbol{P}$, which converts the FOD SH coefficients into FOD amplitudes:

\begin{equation} \label{eq:1a}
    \boldsymbol{u} = \boldsymbol{P} \boldsymbol{f}.
\end{equation}

Then, a regularization matrix $\boldsymbol{L}$ is constructed as follows:

\begin{equation} \label{eq:2a}
L_{m,n} = \begin{cases}
      P_{m,n}, & \text{if $u_m < \tau$ }\\
      0, & \text{if $u_m \geq \tau$,}
    \end{cases} 
\end{equation}

where $\tau$ is a threshold for the FOD amplitude commonly set at 10$\%$ of the mean amplitude \cite{tournier2007robust}. Afterwards, the following minimization problem is solved iteratively until it converges:

\begin{equation} \label{eq:3a}
\boldsymbol{f}_{i+1}
= \operatorname*{arg\,min}_{\boldsymbol{f}_{i+1}}\;
\Big\{
  \lVert \boldsymbol A \boldsymbol{f}_{i+1} - \boldsymbol s \rVert_2^2
  + \lambda\, \lVert \boldsymbol L_i\, \boldsymbol{f}_{i+1} \rVert_2^2
\Big\} ,
\end{equation}

where the first term is the mean squared error of the data fitting, $\boldsymbol{s}$ denotes the $n \times 1$ vector of dMRI data and $\boldsymbol{A}$ is the $n \times p$ matrix defining the inverse problem. The second term is the Tikhonov regularization part, which penalizes FOD amplitudes smaller than the predefined threshold $\tau$, including negative values. The non-negative regularization parameter $\lambda$ defines the trade-off between the data fitting and regularization terms, and $\boldsymbol{L}_i$ is updated at each iteration according to Eqs. (\ref{eq:1a})-(\ref{eq:2a}) using the coefficients vector $\boldsymbol{f}_i$ computed at iteration $i$. Equation (\ref{eq:3a}) is solved in each iteration as:

\begin{equation} \label{eq:4a}
\boldsymbol{f}_{i+1}= (\boldsymbol{A}^T\boldsymbol{A} + \lambda \boldsymbol{L}_i^T\boldsymbol{L}_i)^{-1} \boldsymbol{A}^T\boldsymbol{s}.
\end{equation}

The order $l_{max}$ required to build the matrix $\boldsymbol{A}$ is typically chosen such that the number of SH coefficients $p=(l_{max}+1)(l_{max}+2)/2$ to be estimated is less than or equal to the number of measurements $n$. However, due to the prior information introduced by the non-negativity constraint, it is possible to estimate a number of SH coefficients $p>n$, allowing the FOD to be represented using higher-order SH coefficients. This variant is known as Super-CSD \cite{tournier2007robust}. Despite the superior performance of Super-CSD in resolving narrower fiber crossings, this variation is more sensitive to noise and produces less stable estimates. Therefore, $l_{max}=8$ is used in practice to balance the trade-off between CSD angular resolution and noise robustness.

We noted that although the original SSST-CSD method was implemented using this approach \cite{tournier2007robust}, the implementation in MRtrix3 \cite{MRtrix3} was slightly modified to include a new regularization term to improve the stability of the estimates:

\begin{equation} \label{eq:5a}
\boldsymbol{f}_{i+1}= (\boldsymbol{A}^T\boldsymbol{A} + \lambda_1 \boldsymbol{L}_i^T\boldsymbol{L}_i + \lambda_2 \boldsymbol{I})^{-1} \boldsymbol{A}^T\boldsymbol{s}.
\end{equation}

On the other hand, in the multi-shell multi-tissue CSD (MSMT-CSD) version \cite{Jeurissen2014}, the cost function in Eq. (\ref{eq:3a}) is replaced by

\begin{equation}\label{eq:6}
\begin{split}
\widetilde{\boldsymbol f}
= \operatorname*{arg\,min}_{\boldsymbol f}\;
\big\lVert \boldsymbol A \boldsymbol f - \boldsymbol s \big\rVert_2^2
\quad\\
\text{subject to } \boldsymbol P \boldsymbol f \ge \boldsymbol 0,
\end{split}
\end{equation}

which can be written as a Quadratic Programming (QP) problem, strictly enforcing the non-negativity constraint on the FOD amplitudes:

\begin{equation}\label{eq:7}
\begin{split}
\widetilde{\boldsymbol f}
&= \operatorname*{arg\,min}_{\boldsymbol f}\;
   \Big\{ \tfrac{1}{2}\,\boldsymbol f^{\mathsf T}\boldsymbol Q \boldsymbol f
         + \boldsymbol c^{\mathsf T}\boldsymbol f \Big\} \\
&\quad \text{subject to } \boldsymbol K \boldsymbol f \le \boldsymbol q,
\end{split}
\end{equation}

where $\boldsymbol{K}$=$-\boldsymbol{P}$, $\boldsymbol{q}=\boldsymbol{0}$, $\boldsymbol{Q}$=$\boldsymbol{A}^T \boldsymbol{A}$, and $\boldsymbol{c}^T=-\boldsymbol{s}^T\boldsymbol{A}$.

In this study, the CSD and Super-CSD estimates corresponding to the SSST-CSD method were computed using Eq. (\ref{eq:5a}), employing the default parameters in MRtrix3 \cite{MRtrix3}: $\lambda_1 = k_1 \cdot c$, $\lambda_2 = k_2 \cdot d$, where $c=\max(\boldsymbol{A}^T\boldsymbol{A})$, $d=2 \cdot 10^{-4} \cdot c$, and $k_1$, $k_2$ and $\tau$ are the input parameters with default values of $k_1=k_2=1$ and $\tau=0$. In contrast, the estimates corresponding to MSMT-CSD were computed using Eq. (\ref{eq:7}) utilizing the \textit{quadprog} \cite{quadprog} solver available at \url{https://github.com/qpsolvers/qpsolvers}.

\subsection{Spatially Regularized Super-Resolved CSD (SR$^2$-CSD)}

Nerve fibers should exhibit spatial continuity across WM voxels, as they do not abruptly begin or end their trajectories within the WM. Therefore, utilizing information from estimates in neighboring voxels could lead to more accurate results. The standard CSD formulation does not incorporate spatial information, as it performs the estimation independently for each voxel. This study introduces a novel method, SR$^2$-CSD, which incorporates a spatial prior along with the non-negativity constraint. As demonstrated in our experiments, this additional information helps stabilize the Super-CSD technique, enabling the practical use of larger $l_{max}$ values. This yields estimates with higher angular resolution and fewer spurious FOD lobes compared to standard CSD and Super-CSD.

The SR$^2$-CSD inverse problem incorporates spatial information between adjacent voxel solutions as prior information to enhance denoising and promote spatial continuity of the estimated FODs. It is formulated in the following general form:

\begin{equation}\label{eq:8}
\begin{split}
\widetilde{\boldsymbol f}
&= \operatorname*{arg\,min}_{\boldsymbol f}\;
   \Big\{ \lVert \boldsymbol A \boldsymbol f - \boldsymbol s \rVert_2^2
   + \kappa^2 \, \lVert \boldsymbol f - \boldsymbol f_0 \rVert_2^2 \Big\} \\
&\quad \text{subject to } \boldsymbol P \boldsymbol f \ge \boldsymbol 0 ,
\end{split}
\end{equation}

where $\boldsymbol{f}_0$ denotes a prior estimate of the FOD SH coefficients, computed by incorporating spatial information, and the non-negative parameter $\kappa$ controls the weighting between the data-fitting and regularization terms. Notably, Eq.~(\ref{eq:8}) can be reformulated as a QP problem similar to Eq.~(\ref{eq:7}), but with modified parameters $\boldsymbol{K}=-\boldsymbol{P}$, $\boldsymbol{q}=\boldsymbol{0}$, $\boldsymbol{Q}=\boldsymbol{A}^\top \boldsymbol{A} + \kappa^2\boldsymbol{I}$, and $\boldsymbol{c}^\top=-\boldsymbol{s}^\top\boldsymbol{A}-\kappa^2\boldsymbol{f}_0^\top$. Through empirical analysis, we observed that parameterizing the regularization weight as $\kappa^2=\rho^2 c$, with $c$ as defined in the previous section and setting $\rho=1$, performs robustly across different datasets and noise levels. This parameterization produces data-fidelity and prior terms of comparable magnitude in Eq.~(\ref{eq:8}), like in the standard formulation (see the regularization parameters defined for Eq. (\ref{eq:5a})) .

Various strategies could be implemented to obtain $\boldsymbol{f}_0$. In this study, we computed it as a Total Variation (TV) denoised version of the solution produced by the Super-CSD technique. TV denoising is a highly effective spatial filter that simultaneously removes noise in flat regions while preserving edges. Importantly, the smoothing is primarily applied adaptively along the dominant directions of the image structures, preserving fine details. This property is particularly advantageous for CSD, as it promotes smoothness and continuity along individual tracts, while reducing partial volume contamination between adjacent tracts \cite{Canales2015_rumba}. Note that this formulation can be used in both SSST-CSD \cite{tournier2007robust} and MSMT-CSD \cite{Jeurissen2014} variants.

In practice, the TV filter \cite{rudin1992nonlinear} is applied independently to each 3D image formed by the FOD SH coefficients of a given degree and order. The strength of the filtering, defined by the hyperparameter $\lambda_{TV}=K \sigma$, varies depending on the noise standard deviation $\sigma$ affecting each SH 3D map. This noise standard deviation is estimated using a robust wavelet-based estimator \cite{Donoho1994}, implemented in the \textit{estimate\_sigma} function available in the \textit{scikit-image} Python package \cite{van2014scikit}. In the results section, we demonstrate the impact of $K$ on the solution, which is kept constant for all SH coefficients. However, since the optimal $K$ likely depends on the complexity of the underlying anatomy, it is generally not feasible to define a constant $K$ value that is optimal across different datasets and phantoms. To address this, we employ the J-invariance principle \cite{batson2019noise2self, kobayashi2020image} (see Section \ref{J-inv}) to automatically calibrate the parameter $K$ for each dMRI dataset. In our implementation, TV denoising is implemented using the Chambolle method \cite{chambolle2004algorithm}, available in the \textit{scikit-image} package. The smoothed FOD SH coefficients are then projected onto the constrained non-negative set, ensuring that the resulting SH coefficients produce non-negative FOD amplitudes. This projection involves calculating the FOD amplitudes, setting negative values to zero, and recomputing the SH coefficients from the thresholded FODs. The resulting prior $\boldsymbol{f}_0$ is then used to solve the QP problem in Eq. (\ref{eq:8}). The proposed approach is implemented in the DIPY software \cite{garyfallidis2014dipy}.

\subsection{J-Invariance for Automatic TV Denoiser Calibration}
\label{J-inv}

The automatic calibration of the TV denoiser uses a self-supervised approach based on the Noise2Self algorithm \cite{batson2019noise2self}. A denoising function is considered J-invariant if the value of each pixel in the denoised output does not depend on the value of the same pixel in the noisy input image. If a function is not J-invariant, a masking procedure can be applied to enforce J-invariance. For a J-invariant denoiser, the self-supervised loss approximates the ground-truth loss up to an additive constant \cite{batson2019noise2self}. This implies that the difference between the denoised and noisy images can serve as a proxy for the difference from the original unknown clean image. Therefore, a J-invariant denoiser can thus be calibrated using a self-supervised loss, even in the absence of clean reference images. In practice, we employed the \textit{calibrate$\_$denoiser} function from the \textit{scikit-image} library, which provides an automatic J-invariant calibration of a denoising function by optimizing its parameters through the self-supervised loss. In our experiments, we searched for the optimal TV filtering strength $K$ within the range $[0, 5]$.

\section{Materials and Methods}
The performance of CSD, Super-CSD, and SR$^2$-CSD is assessed using a diverse collection of datasets, comprising two numerical phantoms and two real brain dMRI datasets. Specifically, we employed the HARDI phantom, the DisCo dataset, the single-subject Sherbrooke dataset, and the multi-subject test–retest Stanford dataset. Whereas the HARDI phantom and real brain data were used to evaluate intra-voxel FOD estimation, the DisCo phantom was specifically employed to assess the ability of the methods to recover inter-regional structural connectivity from the estimated FODs, providing ground-truth information that is not available in real brain dMRI acquisitions.

Accordingly, the results are presented in the following order: HARDI phantom, real brain data, and finally the DisCo phantom. The following subsections describe the employed datasets, acquisition parameters, preprocessing steps, and evaluation metrics.

\subsection{HARDI Phantom}
The synthetic dMRI phantom from the \textit{HARDI Reconstruction Challenge 2013}, organized as part of the IEEE International Symposium on Biomedical Imaging 2013, was used to evaluate the algorithms. The dataset is designed to simulate HARDI data by employing a dMRI model that accounts for intra- and extra-cellular water. The phantom consists of 27 fiber bundles connecting different regions of a 3D image with dimensions of $50\times50\times50$ voxels. It features various configurations, including branching, crossing, and kissing, along with diverse fiber bundle radii and geometric properties. Additionally, the phantom incorporates isotropic compartments to mimic CSF partial-volume contamination observed near ventricles in real brain images \cite{daducci2013hardi}. For single-shell single-tissue analysis, the dMRI data were simulated using $n$=64  diffusion directions with a b-value of 3000 s/mm$^2$, and one additional $b=0$ s/mm$^2$ image. For multi-shell multi-tissue analysis, we employed three-shell datasets (b=1000, 2000, and 3000 s/mm$^2$) acquired with the same set of 64 diffusion directions. The single-shell data can be downloaded from the official website: \url{http://hardi.epfl.ch/static/events/2013_ISBI/}.

Rician noise was added to the noise-free dMRI signals $E$ as follows:

\begin{equation} \label{eq:1}
    E_{noisy}=\sqrt{(E+\eta_1)^2 + \eta_2 ^2},
\end{equation}

where $\eta_1$ and $\eta_2$ denote noise vectors sampled from zero-mean, independent Gaussian distributions with variance $\sigma^2$. Noise was added at three signal-to-noise ratio (SNR) levels: 10, 15, and 30, defined as $\text{SNR} = S_0 / \sigma$, with $S_0$ representing the non-diffusion-weighted image for $b=0$ s/mm$^2$. For each SNR level, 10 synthetic datasets were generated using different noise realizations to assess the stability of the methods.

\subsection{Sherbrooke Data}
A multi-shell HARDI dataset was acquired from a healthy human volunteer by the Sherbrooke Connectivity Imaging Lab (SCIL). The dataset includes dMRI images at three b-values (1000, 2000, and 3500 s/mm$^2$), each with 64 diffusion gradient directions. In addition, one non-diffusion-weighted ($b=0$) image was acquired. The spatial resolution was $2 \times 2 \times 2$ mm$^3$. The dataset is publicly available as part of the DIPY software package at \url{https://docs.dipy.org/stable/user_guide/data.html}.

For analysis, the single-shell single-tissue CSD method employed the subset of data with $b$=3500 s/mm$^2$. All three b-values were used to estimate the multi-shell multi-tissue CSD.

\subsection{Test-Retest dMRI Data}
\label{subsec:test-retest}
We assessed the reproducibility of the methods using data acquired at the Center for Cognitive and Neurobiological Imaging at Stanford University, using a 3T GE Discovery MR750 MRI system \cite{rokem2015evaluating}. The dataset consists of scans from six healthy male participants, each undergoing two consecutive sessions. Images were acquired with a spatial resolution of 1.5$\times$1.5$\times$1.5 mm$^3$. The dataset includes 10 $b=0$ images and 96 diffusion gradient directions with a b-value of 2000 s/mm$^2$.

To assess test-retest reproducibility under conditions more representative of clinical and research studies, we subsampled the data to retain 45 directions out of the original 96. Subsampling was performed using a greedy algorithm to ensure that the selected directions were as uniformly distributed as possible over the sphere.

\subsection{Diffusion-Simulated Connectivity (DiSCo) dataset}
We employed the Diffusion-Simulated Connectivity (DiSCo) dataset \cite{Rafael-Patino2021, girard2023tractography}, a synthetic dMRI phantom generated via Monte Carlo simulations, which provides realistic substrates with known ground-truth connectivity. Each phantom consists of approximately twelve thousand synthetic tubular fibers, with diameters ranging from 1.4 µm to 4.2 µm, interconnecting sixteen surface-defined regions of interest (ROIs) in complex configurations that include fiber crossing, kissing, branching, and variable packing densities.

Monte Carlo simulations of dMRI signals were performed using these substrates across four $b$-values  ($b$ = 1000, 1925, 3094, and 13191 s/mm\textsuperscript{2}), with a total of 360 diffusion-weighted volumes and four $b = 0$ images. This dataset is particularly well suited for tractography evaluation, as the known ground-truth connectivity allows for assessing reconstruction accuracy under realistic fiber geometries.

\begin{sloppypar}
For tractography, we used the full 4-shell acquisition and used data at four SNR levels (10, 20, 30, and 50). Streamline reconstruction was performed using the \textit{deterministic$\_$tracking} function available in the DIPY library, employing default tracking parameters with a step size of 0.2 mm and a maximum turning angle of 20°. Seed points were uniformly distributed within the mask at a density of two seeds per dimension per voxel, and tracking was constrained by a binary stopping criterion defined on the same mask.
\end{sloppypar}

\subsection{Evaluation Metrics}
\label{subsec:metrics}

This subsection introduces the four voxel-wise evaluation metrics used in our study, together with the statistical analyses performed to assess differences between methods. Reported values were averaged across the extracted white matter voxels. We deliberately adopted evaluation metrics widely used in the literature (e.g., \cite{daducci2013quantitative}) to simplify and standardize the comparison of our results with previously reported findings.

\begin{itemize}
\item \textit{Mean Squared Error (MSE)}: The FOD SH coefficients estimated from two scans are normalized voxel-wise by their total "energy" using Eq.~(\ref{eq:2}) to eliminate scale differences and spatial changes that may arise in real data (e.g., due to scanner gain drifts, $B_0$ inhomogeneities, or session-to-session variability):

\begin{equation} \label{eq:2}
    SH_{norm}[j] = \frac{SH[j]}{\sqrt{\sum_{i=1}^p SH[i]^2}},
\end{equation}

where the index $j=(l,m)$ encodes the degree and order of the SH coefficients. The MSE between the normalized coefficients $SH_{est1}$ and $SH_{est2}$ is then computed as:

\begin{equation} \label{eq:3}
    MSE = \frac{1}{p} \sum_{j=1}^p(SH_{est1}[j]-SH_{est2}[j])^2.
\end{equation}

\item \textit{Angular Correlation Coefficient (ACC)} is calculated per voxel as in \cite{nath2018shard}:

\begin{equation} \label{eq:4}
    ACC = \frac{\sum_{j=1} ^p SH_{est1}[j] SH_{est2}[j]^*}{\sqrt{\sum_{j=1} ^p |SH_{est1}[j]|^2} \sqrt{\sum_{j=1} ^p |SH_{est2}[j]|^2}},
\end{equation}

where $SH_{est2}[j]^*$ denotes the complex conjugate of $SH_{est2}[j]$.

\item \textit{Angular Error (AE)} is defined as the angular distance (in degrees) between paired fiber orientations $\vv{d}_{est1}$ and $\vv{d}_{est2}$ extracted from each FOD:

\begin{equation} \label{eq:5}
AE = \arccos(\vv{d}_{est1} \cdot \vv{d}_{est2}) \cdot \frac{180}{\pi}.
\end{equation}

Estimated and ground-truth peaks are paired using a greedy matching strategy that minimizes angular discrepancies, and the reported AE corresponds to the mean value across all fibers within each voxel \cite{tuch2002high, Canales2008_peaks}. When the number of estimated and true peaks differs, unmatched directions are also taken into account: each extra peak (either in the estimate or in the ground truth) is assigned the minimal angular distance relative to the available directions in the other set. For consistency and comparability with previous studies, this metric was computed using the evaluation script provided in the \textit{HARDI Reconstruction Challenge 2013}.

\item \textit{Peak Number Error (PNE)} quantifies the normalized discrepancy between the number of resolved peaks:

\begin{equation} \label{eq:6}
PNE = \frac{|M_{est1} - M_{est2}|}{M_{est1}},
\end{equation}

where $M_{est1}$ and $M_{est2}$ are the numbers of peaks in the first and second FODs, respectively. Only peaks above 30\% of the maximum amplitude are considered. A tolerance cone of 20$^\circ$ is used to define matching peaks \cite{Alexander2005, acqua2010, daducci2013quantitative}.
\end{itemize}

For the HARDI phantom, where ground-truth fiber directions are available, $\vv{d}_{est1}$ and $M_{est1}$ are replaced with the true values $\vv{d}_{true}$ and $M_{true}$. In real data, all metrics are computed between FODs from the two scan sessions. Since AE and PNE are not symmetric, their reported values are averaged by alternating the reference FOD.

To evaluate statistical significance across methods, we used a within-subject repeated-measures ANOVA across noise realizations, followed by paired-sample t-tests. Bonferroni correction was applied to adjust for multiple comparisons.

\subsection{Preprocessing}
Before computing the FODs, all dMRI datasets used in this study were denoised using the state-of-the-art Marchenko–Pastur Principal Component Analysis (MPPCA) technique \cite{manjon2013diffusion, veraart2016diffusion, mosso2022mp} as implemented in DIPY.

The Sherbrooke dataset underwent preprocessing to correct for head motion and eddy-current-induced distortions. The Test-Retest dataset had already been preprocessed for motion, and B0 field measurements were used to mitigate EPI-related spatial distortions, as described in the Stanford Digital Repository \url{https://purl.stanford.edu/rt034xr8593}. Despite this preprocessing, residual distortions affected each scan differently, which prevented optimal registration using linear techniques. Since no $b=0$ images with reversed phase-encoding were acquired, advanced distortion correction tools such as FSL’s topup \cite{andersson2003correct, smith2004advances} and eddy \cite{eddy2016} could not be employed. Instead, to improve alignment, the estimated FODs from each scan were non-linearly registered using the symmetric FOD-based algorithm proposed in \cite{raffelt2011symmetric}.
\begin{sloppypar}
Response functions for SSST-CSD and MSMT-CSD were estimated using DIPY. For SSST-CSD, the single-fiber response was estimated from the 200 voxels with the highest fractional anisotropy using the \textit{response\_from\_mask\_ssst} function. 
For MSMT-CSD, tissue-specific response functions (WM, GM, CSF) were obtained using the \textit{mask\_for\_response\_msmt} function with default DIPY parameters: 
\texttt{wm\_fa\_thr=0.7}, \texttt{gm\_fa\_thr=0.3}, \texttt{csf\_fa\_thr=0.15}, \texttt{gm\_md\_thr=0.001}, \texttt{csf\_md\_thr=0.0032} (see \url{https://docs.dipy.org/stable/examples_built/reconstruction/reconst_mcsd.html}).
\end{sloppypar}

\section{Results}
\subsection{HARDI Phantom}

In the HARDI dataset, we compared the AE and PNE metrics for CSD with $l_{max}=8$ (i.e., 45 SH coefficients), and for Super-CSD and SR$^2$-CSD, both using $l_{max}=12$ (i.e., 91 SH coefficients). For SR$^2$-CSD, the prior $\boldsymbol{f}_0$ was computed using the optimal filtering strength $K$ obtained from the J-invariant self-calibrated TV denoising method. Additionally, we evaluated the behavior of the SR$^2$-CSD solution as a function of $K$, exploring a discrete set of predefined values.

\begin{figure}[h!]
\centering 
\includegraphics[width=.95\textwidth]{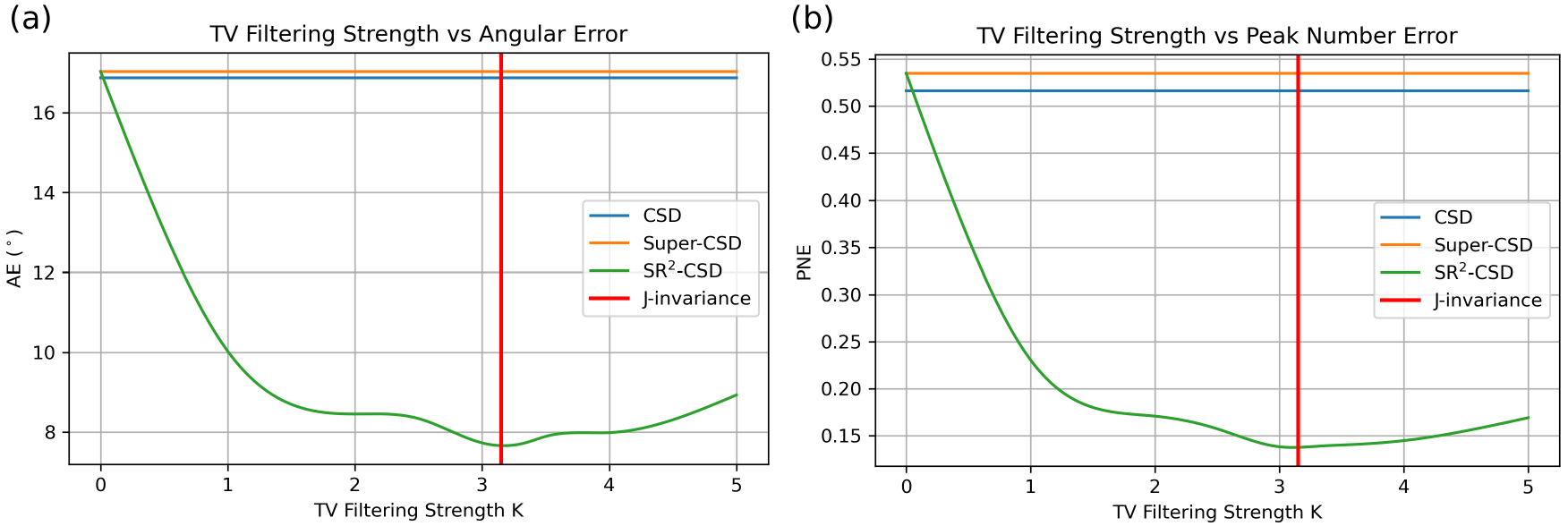}
\caption{Selection of the TV denoising strength $K$ for SR$^2$-CSD with $l_{max} = 12$ in the HARDI Phantom (SNR = 15). (a) Angular Error (AE). (b) Peak Number Error (PNE). Horizontal lines show baseline performance of standard CSD ($l_{max} = 8$, blue) and Super-CSD ($l_{max} = 12$, orange). The vertical red line indicates the $K$ value automatically selected via J-invariance, which matches the minimum of the curves.}
\label{hardi_metrics}
\end{figure}

Figure \ref{hardi_metrics} shows how AE and PNE vary with $K$ in SR$^2$-CSD, for the single-shell single-tissue configuration at SNR = 15 with denoised data. The vertical red line marks the $K$ value automatically selected by the J-invariant calibration, which closely aligns with the global minimum of both AE and PNE. Within the tested $K$ range, any non-zero value of $K$ yielded improved performance over both CSD and Super-CSD (shown for reference as blue and orange horizontal lines, respectively).

\begin{figure}[h!]
\centering 
\includegraphics[width=.9\textwidth]{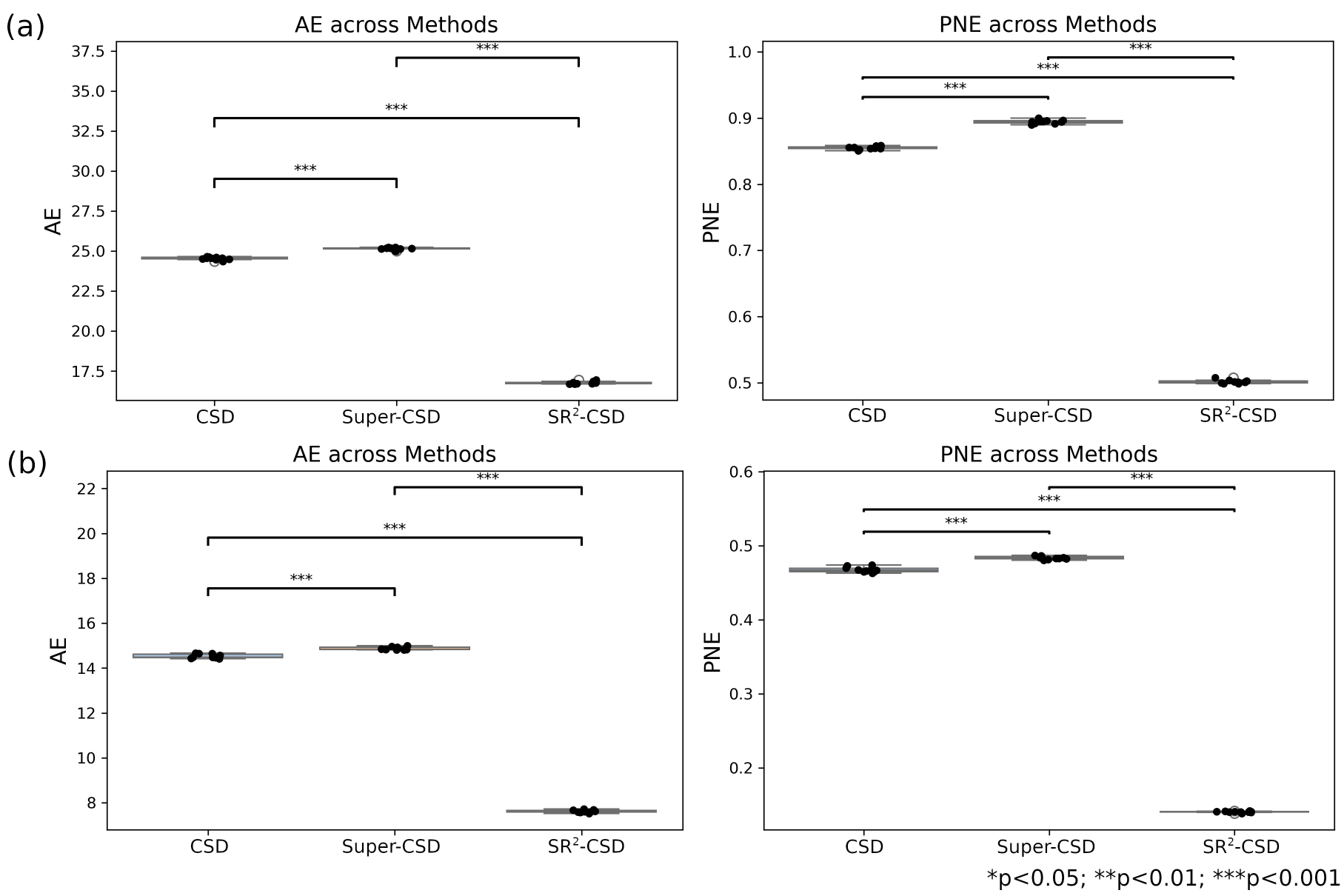}
\vspace{-1em}
\caption{Reconstruction accuracy on the HARDI phantom (SNR = 30, 10 noise realizations, for single-shell single-tissue reconstruction). (a) Without MPPCA denoising. (b) After MPPCA denoising. Boxplots show Angular Error (AE) and Peak Number Error (PNE) for CSD ($l_{max}=8$), Super-CSD ($l_{max}=12$), and SR$^2$-CSD ($l_{max}=12$). Statistical significance assessed via paired t-tests with Bonferroni correction. Asterisks indicate significance levels: *$p < 0.05$; **$p < 0.01$; ***$p < 0.001$.}
\label{hardi_ssst_boxplot_snr30}
\end{figure}

\begin{figure}[h!]
\centering 
\includegraphics[width=.9\textwidth]{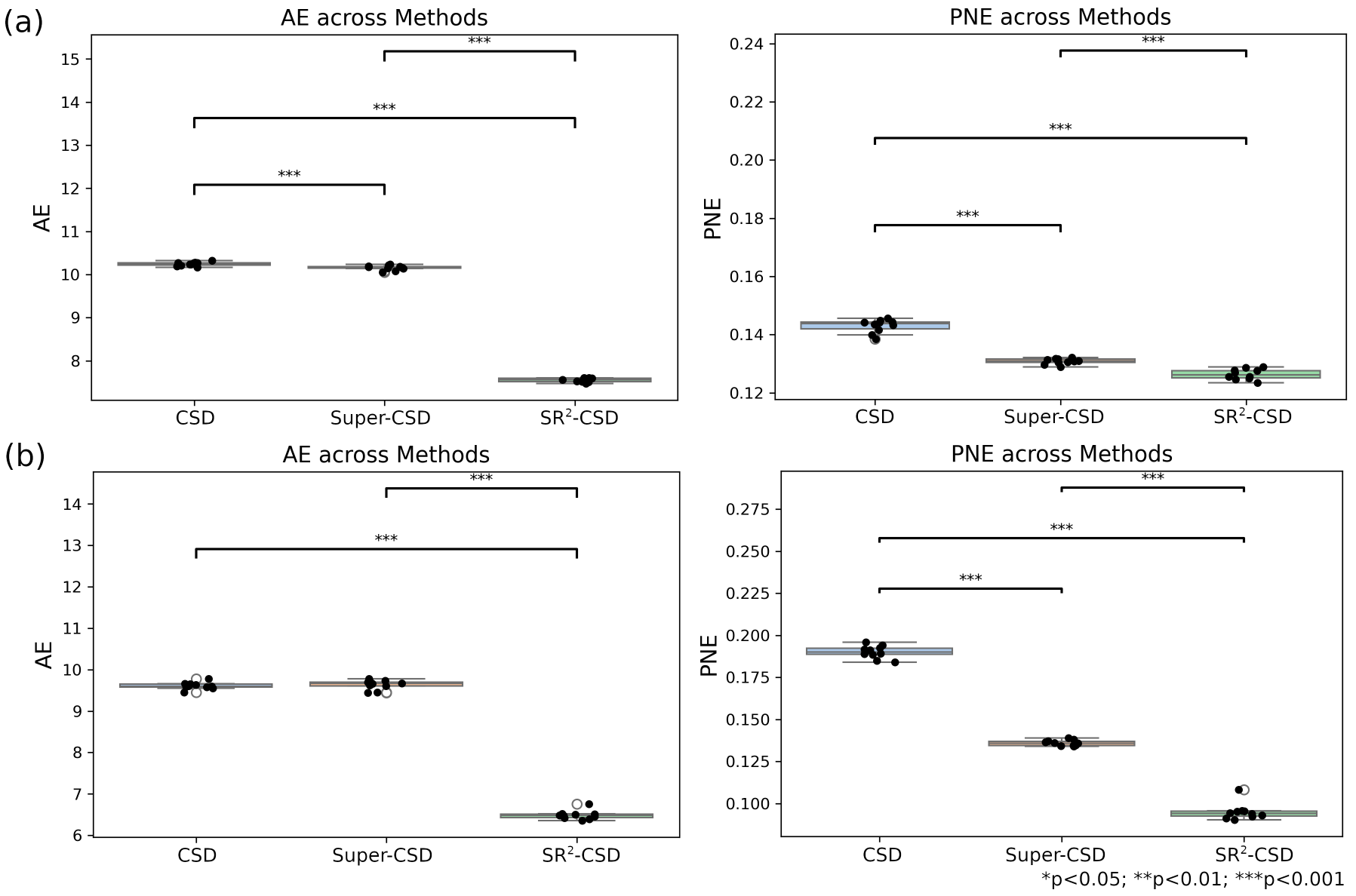}
\vspace{-1em}
\caption{Reconstruction accuracy on the HARDI phantom (SNR = 30, 10 noise realizations, for multi-shell multi-tissue reconstruction). (a) Without MPPCA denoising. (b) After MPPCA denoising. Boxplots show Angular Error (AE) and Peak Number Error (PNE) for CSD ($l_{max}=8$), Super-CSD ($l_{max}=12$), and SR$^2$-CSD ($l_{max}=12$). Statistical significance assessed via paired t-tests with Bonferroni correction. Asterisks indicate significance levels: *$p < 0.05$; **$p < 0.01$; ***$p < 0.001$.}
\label{hardi_msmt_boxplot_snr30}
\end{figure}

\begin{figure}[h!]
\centering 
\includegraphics[width=0.95\textwidth]{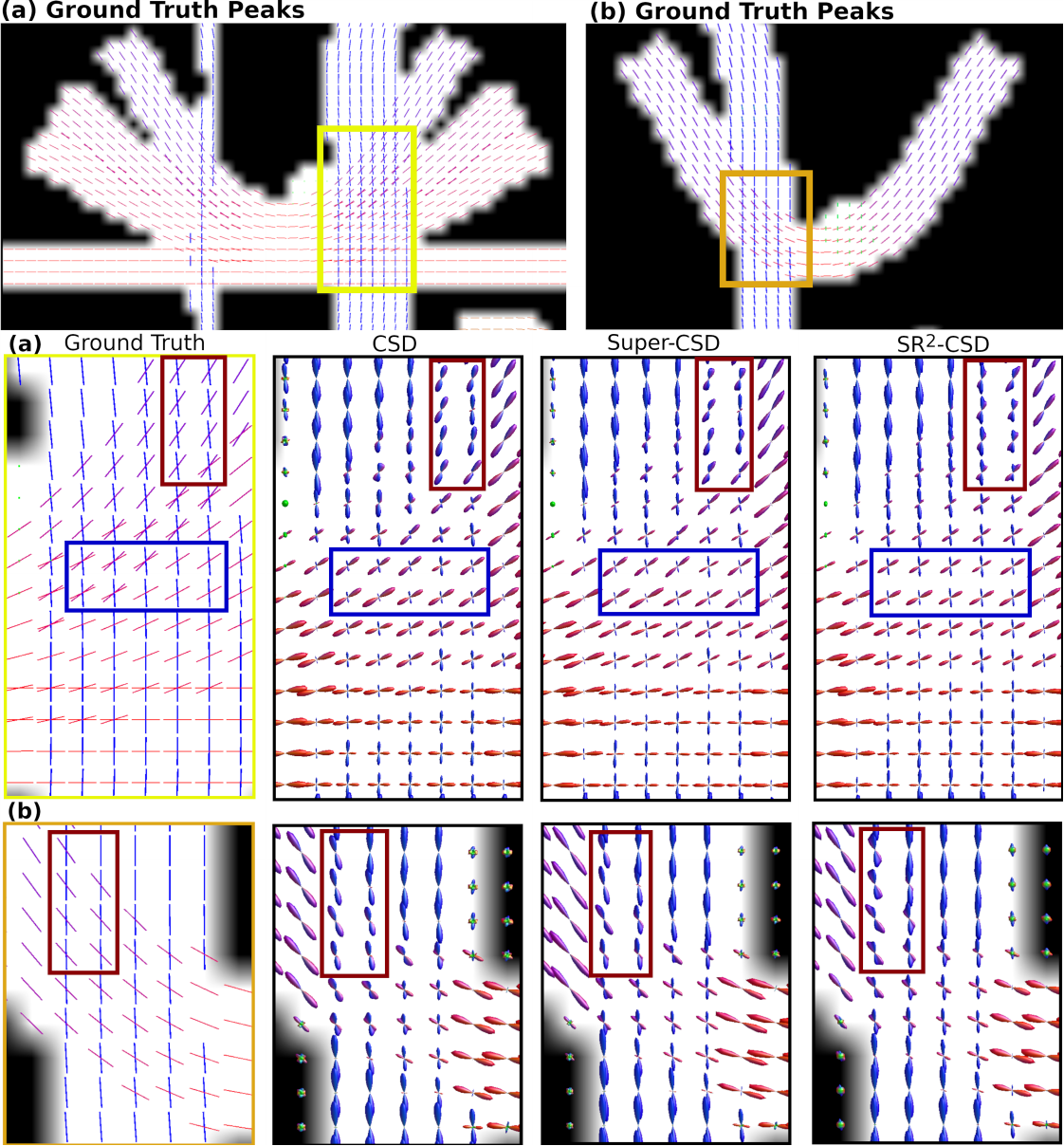}
\caption{Visual comparison of estimated Fiber Orientation Distributions (FODs) in two regions of the HARDI Phantom with fiber crossings. Data: single-shell ($b=3000;s/\text{mm}^2$), SNR = 15, after MPPCA denoising. Row 1: ground-truth fiber orientations. Rows 2–3: estimated FODs from CSD ($l_{max}=8$), Super-CSD ($l_{max}=12$), and SR$^2$-CSD ($l_{max}=12$). Three regions are highlighted where SR$^2$-CSD improved the angular resolution (red boxes) and the angular coherency (blue box).}
\label{hardi_fig}
\end{figure}

A summary of the AE and PNE metrics under 10 different noise realizations at SNR = 30 is presented in Figures \ref{hardi_ssst_boxplot_snr30} and \ref{hardi_msmt_boxplot_snr30}, corresponding to the single-shell single-tissue and multi-shell multi-tissue analyses, respectively. In both cases, results are presented with (panel b) and without (panel a) MPPCA denoising. The tight spread of the boxplots indicates low variability across noise realizations, confirming the robustness of each method. Comparable results for SNR = 10 and 15 are provided in Figures S1–S4 of the Supplementary Material. Across all tested SNR levels and preprocessing conditions, SR$^2$-CSD significantly outperformed both CSD and Super-CSD in terms of AE and PNE, with differences being statistically significant. Summary tables S1–S4 in the Supplementary Material report mean and standard deviation values for all analyses.

To complement these quantitative results, Figure \ref{hardi_fig} provides a visual comparison of FODs estimated in various regions of interest from the MPPCA-denoised single-shell dataset at SNR = 15 (one noise realization). In regions with small inter-fiber angles, CSD and Super-CSD fail to resolve some crossings, while SR$^2$-CSD successfully identifies them (see red boxes). Furthermore, SR$^2$-CSD achieves better angular coherence, producing smoother and more consistent FOD lobes that align with the ground truth (see blue box).

\begin{figure}[h!]
\centering
\includegraphics[width=1\textwidth]{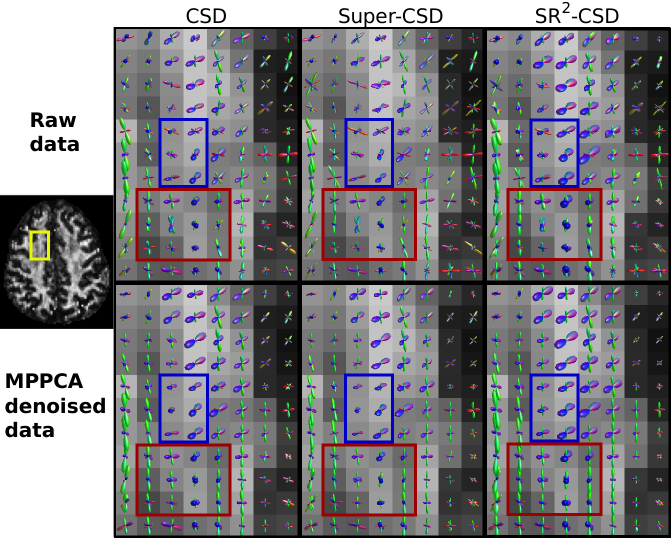}
\caption{Fiber Orientation Distributions (FODs) estimated from the raw Sherbrooke data and MPPCA-denoised data acquired with a single shell ($b = 3500$ s/mm$^2$, 64 gradient directions). From left to right, the figure shows results for CSD with $l_{\text{max}} = 8$, Super-CSD with $l_{\text{max}} = 12$, and SR$^2$-CSD with $l_{\text{max}} = 12$. The red boxes highlight regions where denoising improves the FOD estimates from CSD and Super-CSD, yielding more spatially coherent lobes and reducing isolated peaks that do not appear in neighboring voxels. The blue box marks an adjacent superior region where, even after denoising, the anterior–posterior green lobes visible in the red boxes become markedly attenuated or nearly disappear. In contrast, SR$^2$-CSD preserves these lobes across both regions, especially in the denoised data, producing greater spatial continuity.}
\label{sherbrooke_fig2}
\end{figure}

\begin{figure}[h!]
\centering
\includegraphics[width=0.95\textwidth]{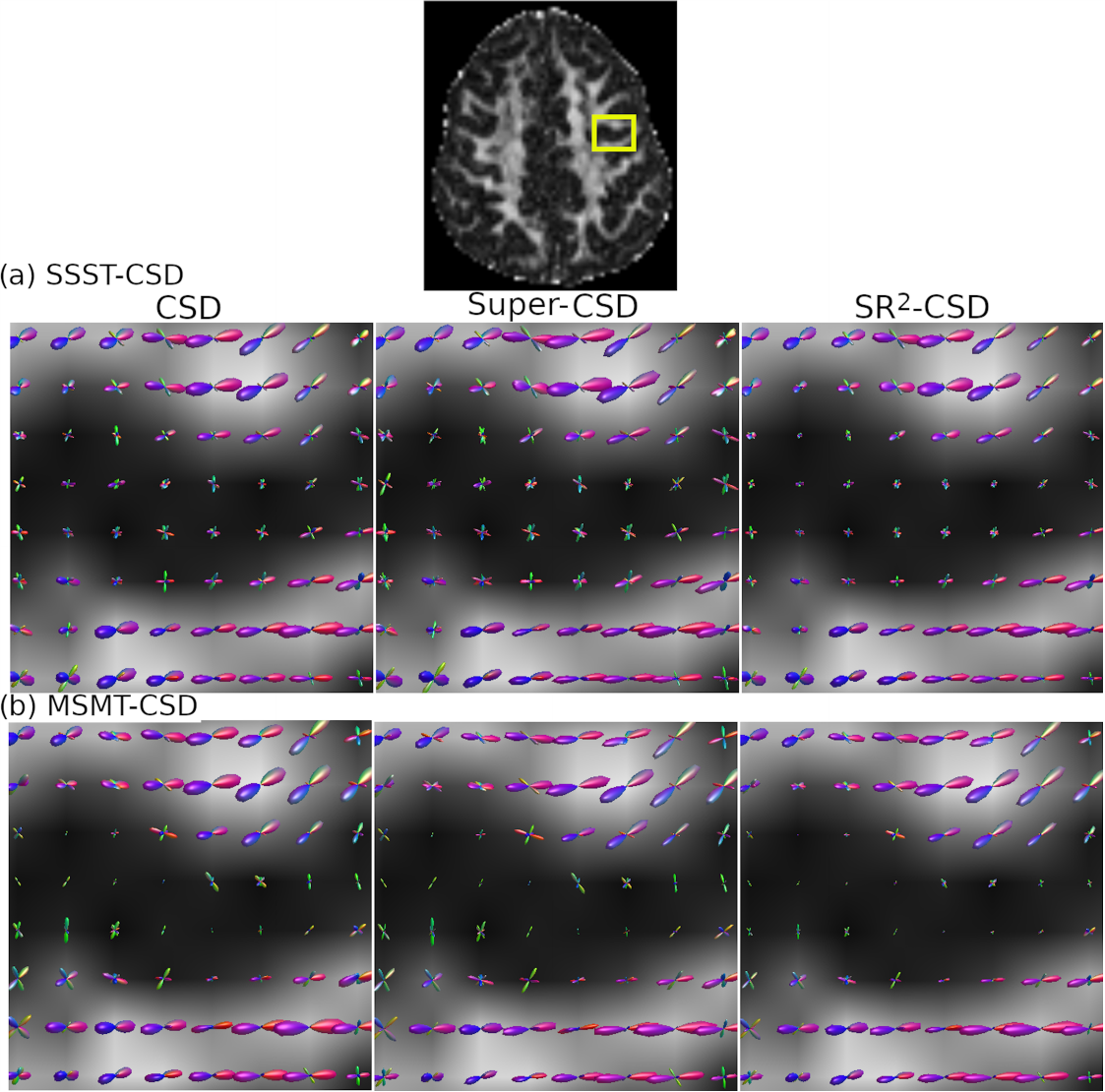}
\caption{Fiber Orientation Distributions (FODs) estimated from the MPPCA-denoised Sherbrooke dataset within a region of interest encompassing white matter (WM), gray matter (GM), and cerebrospinal fluid (CSF) voxels. From left to right, the estimated FODs correspond to: (i) CSD with $l_{max}=8$, (ii) Super-CSD with $l_{max}=12$, and (iii) SR$^2$-CSD with $l_{max}=12$. Results are shown for both the single-shell single-tissue CSD (SSST-CSD, using data with $b$ = 3500 s/mm$^2$) and the multi-shell multi-tissue CSD (MSMT-CSD, using three b-values: 1000, 2000, and 3500 s/mm$^2$).}
\label{sherbrooke_wm_csf}
\end{figure}

\begin{figure}[h!]
\centering
\includegraphics[width=.96\textwidth]{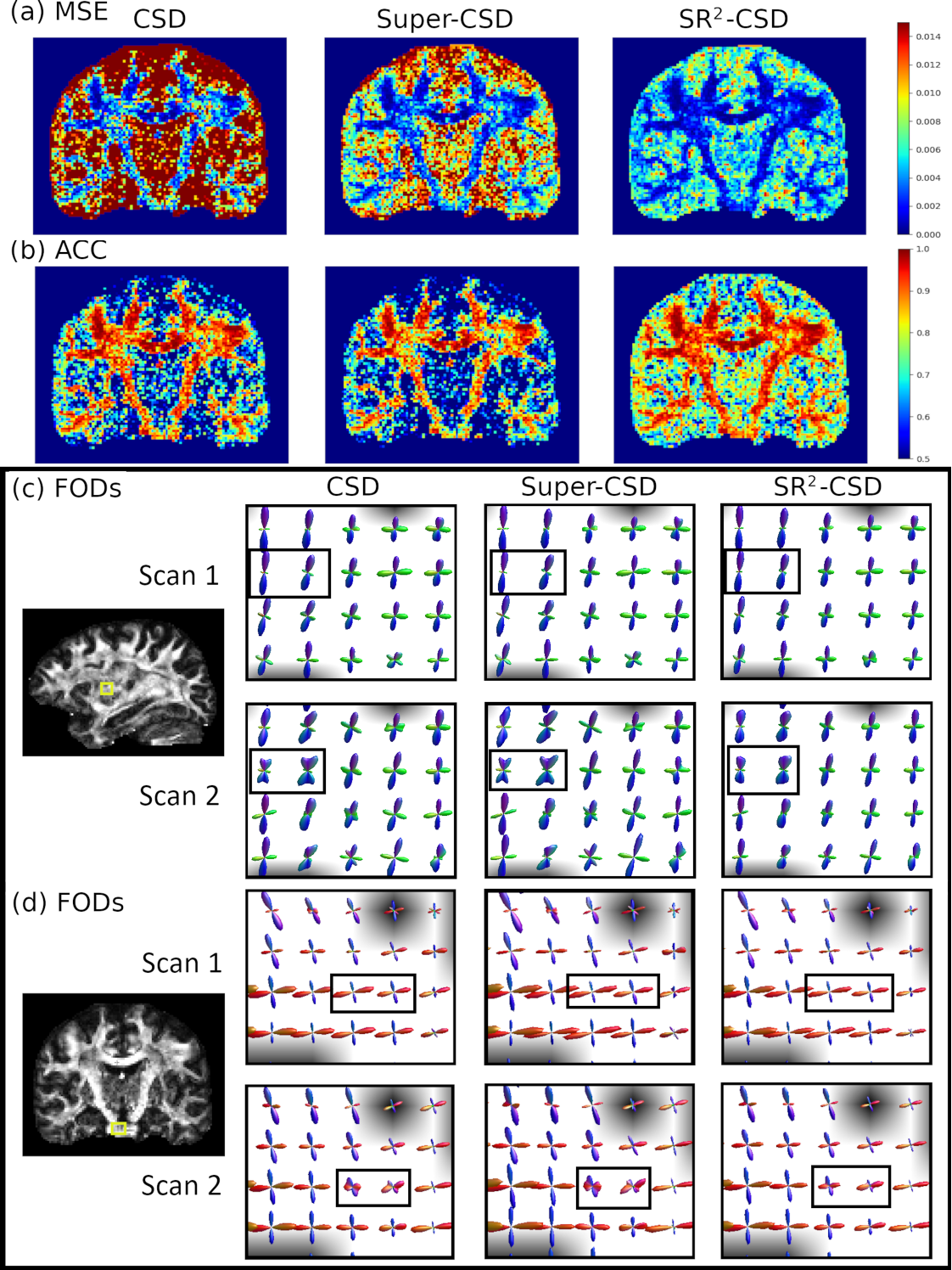}
\caption{Scan-rescan results for Subject 3 using MPPCA-denoised data acquired with the full diffusion scheme (96 gradient directions). Panels (a)-(b) display voxel-wise maps of Mean Squared Error (MSE) and Angular Correlation Coefficient (ACC) computed from FODs estimated using CSD ($l_{max} = 8$), Super-CSD ($l_{max} = 12$), and SR$^2$-CSD ($l_{max} = 12$). Panels (c)-(d) show the corresponding FODs estimated from the two scan sessions in two regions of interest. Black boxes highlight areas where SR$^2$-CSD yielded more consistent FODs across sessions.}
\label{mse_acc_maps}
\end{figure}

\subsection{Sherbrooke Data}
Figure~\ref{sherbrooke_fig2} examines a region of an axial slice located in the superior part of the brain, slightly anterior to the midline and within the left hemisphere, where multiple fiber populations intersect. Based on the spatial context and the orientation-encoded color representation (RGB), we identified two dominant fiber orientations. The blue lobes, oriented inferior–superior with slight obliquity, are consistent with fibers from either the corpus callosum (CC) or the corona radiata. These structures are anatomically adjacent and difficult to distinguish in this location (see, e.g., \cite{catani2008}). The green lobes, oriented anterior–posterior, are consistent with the trajectory of the superior longitudinal fasciculus I (SLF I), which connects the superior frontal and parietal cortices and runs dorsally, as described by \cite{makris2005}. We present results obtained from both the raw noisy data and the MPPCA-denoised data. The FODs estimated by CSD and Super-CSD on the raw data show limited spatial coherence, as illustrated in the blue and red boxes. After denoising, the FODs become more spatially consistent, with improved orientational coherence and fewer isolated peaks that appear in a single voxel but are absent in neighboring ones. This demonstrates the  sensitivity of these methods to noise and shows that, once noise is attenuated, the alignment of neighboring FODs within the same fiber bundle becomes more evident. In contrast, the FODs estimated by SR$^2$-CSD remain more stable in both data. The reconstructions obtained with SR$^2$-CSD are smoother and more continuous, especially in the denoised data. For instance, in the red boxes, the anterior–posterior fiber components (green lobes) estimated by CSD and Super-CSD are markedly attenuated and nearly disappear in the superior adjacent blue boxes, even in the denoised data. In comparison, SR$^2$-CSD preserves these lobes across contiguous voxels without abrupt discontinuities. All results were obtained using the SSST-CSD implementation on data acquired at $b = 3500$ s/mm$^2$. The optimal filtering strength for SR$^2$-CSD was automatically determined via the J-invariant total-variation calibration ($K = 0.45$).

Figure~\ref{sherbrooke_wm_csf} shows representative FODs from both SSST-CSD and MSMT-CSD variants at the WM–GM–CSF interface in an example region of interest. Across all methods, the SSST-CSD reconstructions exhibit increased noise, as expected, given that a single $b$-value lacks sufficient sensitivity to disentangle signal contributions from distinct tissue compartments. Notably, the SR$^2$-CSD reconstructions exhibit reduced angular artifacts and more anatomically plausible FODs, with lower amplitudes outside WM voxels and fewer spurious lobes in regions unlikely to contain fiber bundles.

\subsection{Test-Retest dMRI Data}
To assess reproducibility, we analyzed two FOD reconstructions per subject obtained from two separate scan sessions, for each of the six participants in the dataset. The similarity between FODs was quantified using four metrics introduced in Section~\ref{subsec:metrics}: MSE, ACC, AE, and PNE. Lower values of MSE, AE, and PNE, and higher values of ACC, indicate higher reproducibility. All metrics were computed within a WM mask constructed by selecting voxels with fractional anisotropy (FA) $>$ 0.3. The TV filtering strength estimated automatically via the J-invariant procedure had a mean value of $K$ = 0.44 (standard deviation: 0.08) across all subjects.

Figure~\ref{mse_acc_maps} presents the scan-rescan reproducibility results for a representative subject (Subject 3), based on MPPCA-denoised data acquired with the full scheme of 96 diffusion gradient directions. Voxel-wise MSE and ACC maps are shown in Panels (a) and (b), respectively, for the three methods: CSD with $l_{\text{max}} = 8$, Super-CSD with $l_{\text{max}} = 12$, and SR$^2$-CSD with $l_{\text{max}} = 12$. SR$^2$-CSD yielded the most reproducible FODs, evidenced by lower MSE and higher ACC values. Panels (c) and (d) show the FOD reconstructions from the two sessions for two regions of interest in sagittal and coronal slices. The highlighted voxels illustrate areas where SR$^2$-CSD provided better visual consistency across scans.

To complement these maps, we computed histograms of voxel-wise differences in MSE and ACC between SR$^2$-CSD and the other two methods (Figure S6, Supplementary Material). These histograms were derived from all voxels within the WM mask. SR$^2$-CSD yielded lower MSE values in 99.77\% and 99.59\% of voxels compared to CSD and Super-CSD, respectively, and higher ACC values in 85\% and 99.59\% of voxels. This voxel-wise analysis further confirms the superior reproducibility of the proposed method.

\begin{figure}[h!]
\centering
\includegraphics[width=.92\textwidth]{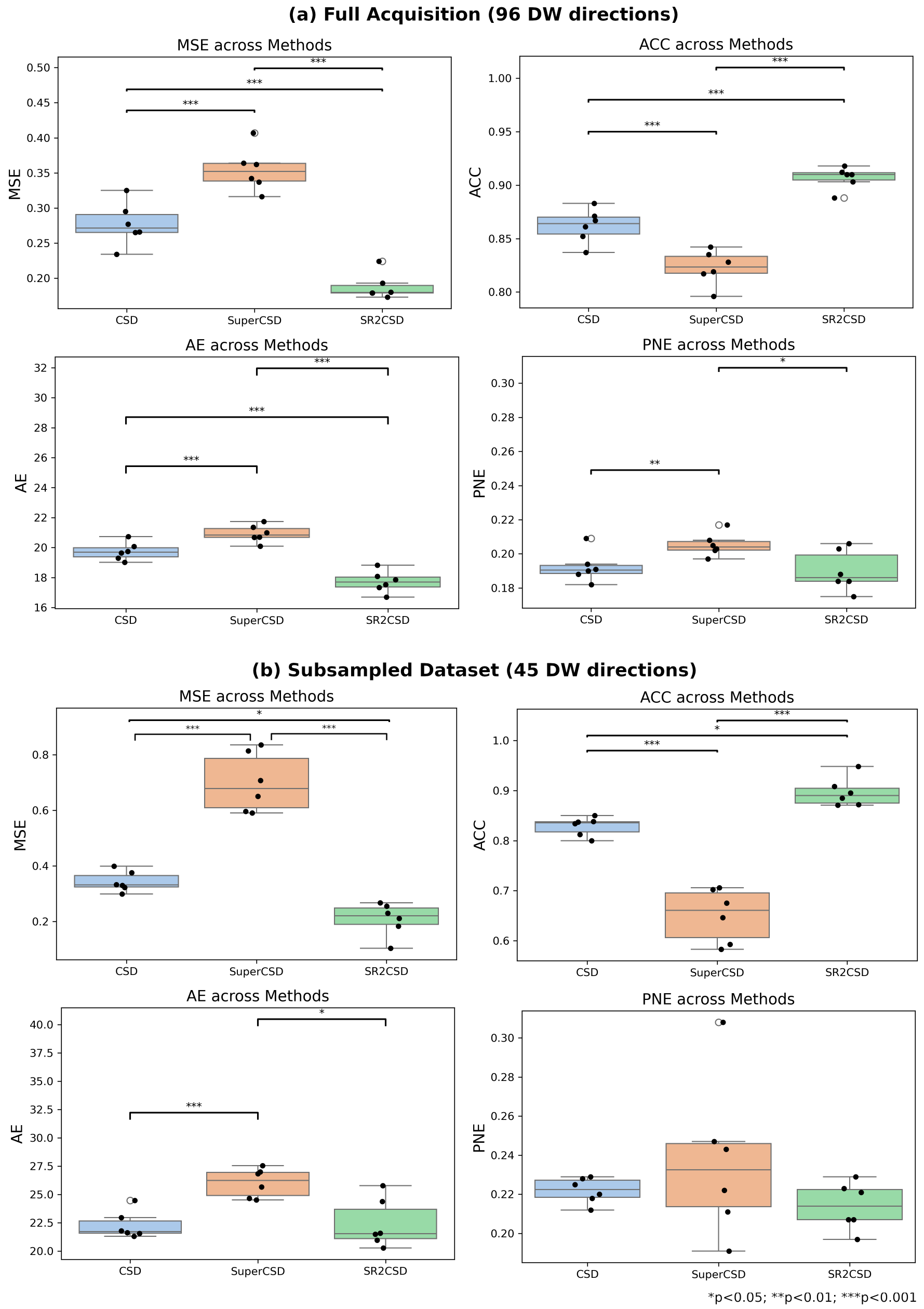}
\caption{Comparison of method performance using the test–retest dataset.
Panel (a) shows results using the full acquisition scheme (96 gradient directions), and panel (b) shows results with the subsampled dataset (45 gradient directions). Metrics were computed from FODs estimated across two scan sessions per subject. Paired-sample t-tests with Bonferroni correction were used to assess the statistical significance of pairwise differences between methods. Asterisks denote significance levels: (*$p<0.05$; **$p<0.01$; ***$p<0.001$).}
\label{96_45_dir_test_retest_boxplots}
\end{figure}

Figure~\ref{96_45_dir_test_retest_boxplots} summarizes the reproducibility metrics across all six subjects, comparing results from both the full acquisition scheme (96 directions) and the subsampled scheme (45 directions). SR$^2$-CSD consistently achieved statistically significant improvements over CSD and Super-CSD for most metrics. Specifically, it produced significantly lower MSE and AE, and higher ACC values across both sampling schemes. PNE was also generally lower for SR$^2$-CSD. In 12 out of the 16 pairwise comparisons (4 metrics × 2 sampling schemes × 2 references), SR$^2$-CSD significantly outperformed both reference methods. Tables S5 and S6 in the Supplementary Material detail the mean values per subject, method, and metric. Across all subjects, SR$^2$-CSD consistently yielded lower MSE and higher ACC than both CSD and Super-CSD. In addition, it achieved the lowest AE in all subjects for the full acquisition and in four subjects for the subsampled case. Similarly, SR$^2$-CSD obtained the lowest PNE in five subjects with the full acquisition and in four subjects with the subsampled protocol.

\subsection{Diffusion-Simulated Connectivity (DiSCo) dataset}
We evaluated tractography accuracy on the DiSCo dataset by comparing the estimated and ground-truth connectivity matrices. The ground-truth matrix was defined from the total cross-sectional area of the synthetic fibers connecting each pair of the sixteen ROIs. For each method and SNR level, a $16 \times 16$ connectivity matrix was computed from the corresponding tractogram. Tractography accuracy was quantified using the Pearson correlation coefficient ($r$) between the estimated and ground-truth matrices, following the official evaluation protocol of the DiSCo Challenge \cite{girard2023tractography}, where this metric was used to rank submitted methods.

Figure~\ref{disco_tract} and Table~S7 in the Supplementary Material summarize the results. Table~S7 reports the mean and standard deviation of the correlation coefficients across four SNR levels, using 10 independent noise realizations for each method (CSD, Super-CSD, and SR$^2$-CSD). Figure~\ref{disco_tract} presents the corresponding boxplots and statistical analyses. Across all conditions, SR$^2$-CSD consistently achieved the highest correlations with the ground-truth connectivity, outperforming both reference methods (see Table~S7). Statistical analyses in Figure~\ref{disco_tract} revealed significant differences between SR$^2$-CSD and both CSD and Super-CSD for SNR = 10, 20, and 30, whereas for SNR = 50 no significant difference was found between SR$^2$-CSD and CSD (only between SR$^2$-CSD and Super-CSD). When comparing CSD and Super-CSD, both yielded statistically similar correlations for SNR = 10, 20, and 50, but differed at SNR = 30, where CSD outperformed Super-CSD.

\begin{figure}[h!]
\centering
\includegraphics[width=1\textwidth]{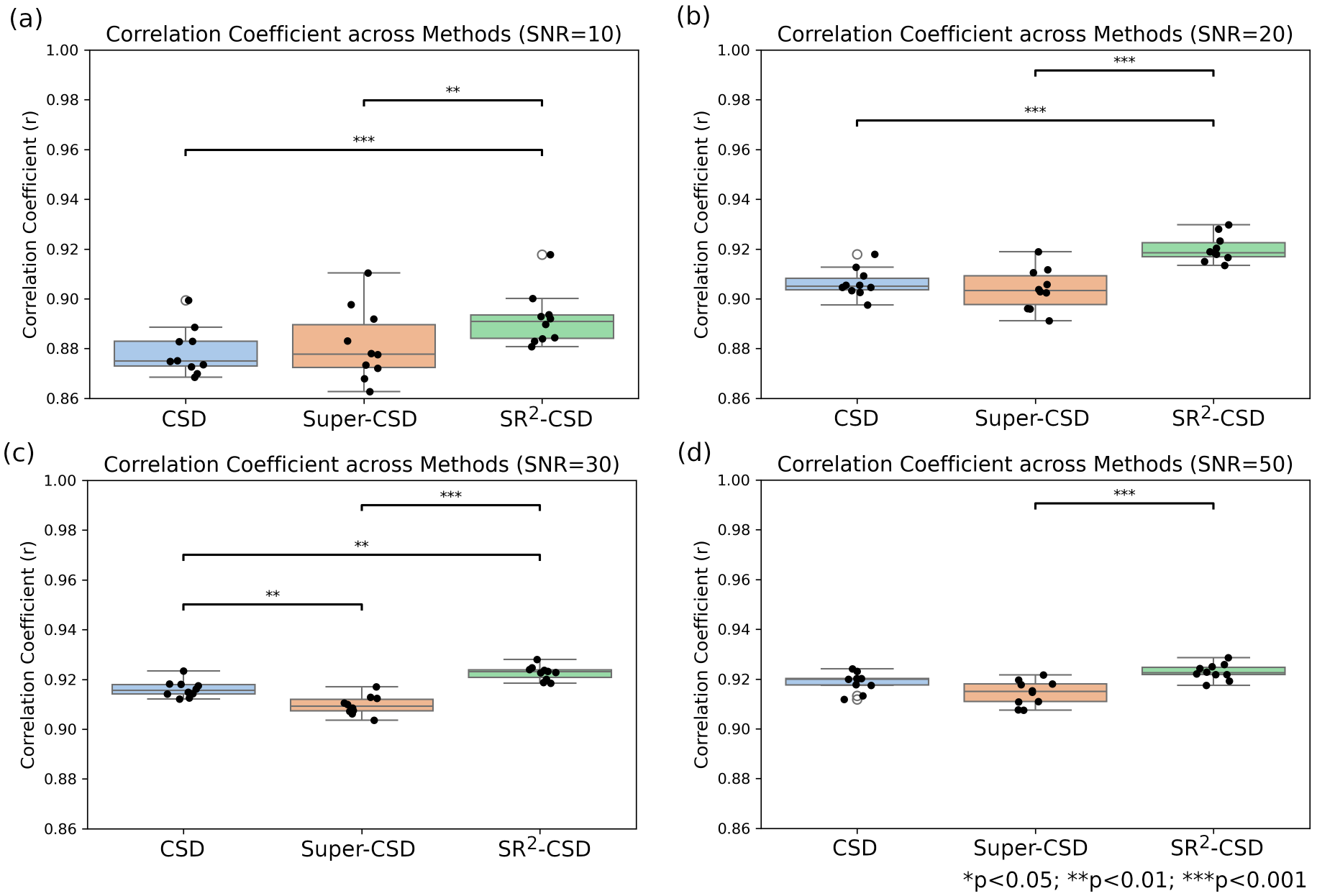}
\vspace{-1em}
\caption{Pairwise statistical comparison of correlation coefficients across reconstruction methods on the DiSCo dataset under varying noise levels. Boxplots illustrate the distribution of correlation coefficients (r) for CSD, Super-CSD, and SR$^2$-CSD at four SNR levels: (a) SNR = 10, (b) SNR = 20, (c) SNR = 30, and (d) SNR = 50 under Rician noise, computed over 10 noise realizations for each method. Paired-sample t-tests with Bonferroni correction were used to assess the statistical significance of pairwise differences between methods. Asterisks denote significance levels: (*$p<0.05$; **$p<0.01$; ***$p<0.001$).}
\label{disco_tract}
\end{figure}

\section{Discussion and Conclusion}

In this proof-of-concept study, we propose a novel Spatially Regularized Super-Resolved CSD (SR$^2$-CSD) method as an extension of the state-of-the-art CSD framework. It leverages a spatial prior obtained through TV filtering of the FOD estimated by the Super-CSD method. The filtering strength is automatically calibrated using the J-invariance principle, a technique grounded in self-supervised blind denoising.

We evaluated the performance of the proposed method against standard CSD with $l_{max}=8$ and Super-CSD with $l_{max}=12$. Our experiments employed two phantoms (the HARDI phantom and the DiSCo dataset) and two real brain datasets (the single-subject Sherbrooke dataset and the multi-subject test–retest Stanford dataset). Performance was comprehensively assessed through quantitative metrics, statistical tests, and qualitative visual inspection. 

The impact of noise on the estimated FODs was examined by corrupting the HARDI phantom data at three different SNR levels, with multiple noise realizations per level. We considered both single-shell single-tissue and multi-shell multi-tissue CSD variants, and evaluated results using raw and MPPCA-denoised dMRI data \cite{manjon2013diffusion, veraart2016diffusion, mosso2022mp}. The intra-voxel metrics and statistical analyses showed that SR$^2$-CSD consistently achieved lower angular and peak number errors across SNR levels, in both single- and multi-shell reconstructions, and with raw or denoised data (Figures \ref{hardi_ssst_boxplot_snr30}–\ref{hardi_msmt_boxplot_snr30}, Figures S1–S4, and Tables S1–S4 of the Supplementary Material). Visual comparisons (Figure \ref{hardi_fig}) corroborated these findings, highlighting the superior ability of SR$^2$-CSD to resolve small inter-fiber angles and to improve spatial-angular coherence of the FODs.

In the Sherbrooke dataset, SR$^2$-CSD produced smoother FOD estimates without sacrificing angular resolution, potentially offering a better representation of continuous fiber tracts (Figure \ref{sherbrooke_fig2}). Single-shell and multi-shell analyses were compared at the WM–GM–CSF interfaces. In line with previous studies \cite{Jeurissen2014}, single-shell data proved insufficient to model contributions from non-WM compartments (Figure \ref{sherbrooke_wm_csf}). Notably, the inherent denoising capability of SR$^2$-CSD reduced the noisy FODs obtained with CSD and Super-CSD in these regions. A similar trend was observed in multi-shell analyses, though all methods produced smoother FODs with fewer spurious lobes.

Reproducibility was assessed using the Stanford test–retest dataset, comprising six subjects scanned twice. To approximate clinical conditions, we evaluated both the full acquisition (96 gradient directions) and a subsampled version (45 gradient directions). Across four similarity metrics, SR$^2$-CSD yielded more reproducible reconstructions in both schemes. Quantitative results (Figure \ref{96_45_dir_test_retest_boxplots} and Tables S5–S6 of the Supplementary Material) were consistent with visual inspection (Figure \ref{mse_acc_maps} and Figure S6 of the Supplementary Material). These analyses demonstrate that SR$^{2}$-CSD improves reproducibility of FOD estimation.

Beyond intra-voxel reconstruction, we conducted a preliminary analysis to test whether the improved voxel-wise FOD estimates provided by SR$^2$-CSD translate into more reliable structural connectivity derived from tractography. Using the DiSCo challenge dataset \cite{girard2023tractography}, we found that connectivity matrices derived from SR$^2$-CSD achieved higher correlations with ground-truth values than those from CSD or Super-CSD, consistently across SNR levels.

It should be emphasized, however, that the correlation coefficients reported here are not directly comparable to those from the original DiSCo challenge \cite{girard2023tractography}, for several reasons. First, all participating teams using MSMT-CSD applied additional streamline post-filtering strategies to reduce false positives, from trajectory- and length-based pruning to advanced global techniques such as SIFT2 \cite{Smith2015} and COMMIT \cite{COMMIT}; indeed, all top-ranked teams employed SIFT2 or COMMIT. In contrast, our metrics were computed using all generated streamlines, since identifying the optimal filtering strategy was beyond the scope of this work. Second, while top-ranked teams used the Parallel Transport Tractography algorithm \cite{Aydogan2021}, we used standard deterministic tracking with default parameters, without fine-tuning for performance, as optimizing and comparing different tracking methods is beyond the scope of this work. Third, challenge participants could submit up to ten connectivity matrices based on different pipelines (e.g., varying filtering, tracking method, tracking parameters, or mask generation), with the best correlation reported. By contrast, we reported results from a single pipeline with default settings, since our goal was to assess the impact of improved FOD estimation rather than optimize the processing pipeline for this dataset.

The computation time of SR$^2$-CSD is higher than that of Super-CSD, as it requires solving the QP problem twice: once for the initial Super-CSD solution used to build the prior, and again for the final SR$^2$-CSD estimation. The second step is faster, however, as the prior accelerates convergence. Using an AMD Ryzen 9 5900X 12-Core processor, computation times were as follows: for the HARDI data—Super-CSD (34 s), TV denoising (32 s), SR$^2$-CSD fitting (22 s); and for the Sherbrooke data—Super-CSD (7 min 36 s), TV denoising (10 min 34 s), SR$^2$-CSD fitting (1 min 28 s). The runtime of the TV filtering step could be substantially reduced by narrowing the explored range of $K$, since the optimal value was found consistently within 0.35–0.65 for all the analyzed real brain datasets, although here we adopted a broad exploratory range (0–5.0).

Another key finding is that all FOD-based quality metrics improved when dMRI data were denoised with the MPPCA method \cite{veraart2016diffusion}, compared to raw data. This step was essential to ensure a fair comparison between CSD, Super-CSD, and SR$^2$-CSD. Without denoising, performance differences could largely be attributed to the extra noise suppression by SR$^2$-CSD. Thus, we applied MPPCA before all reconstructions to enable equitable comparisons. Nevertheless, for completeness, we also reported results on raw data, where SR$^2$-CSD showed even larger relative improvements, as expected.

This is not the first study to apply TV denoising to FOD estimation. In a previous work \cite{Canales2015_rumba}, we incorporated TV filtering into the multiplicative updates of a modified RL-SD algorithm \cite{acqua2007, acqua2010, Deluca, DeLuca2020}, which was also adapted for non-Gaussian MRI noise. Its superior performance motivated our choice of TV here. Importantly, there is a key difference between the RUMBA-SD+TV method implemented in \cite{Canales2015_rumba} and the present study: While RUMBA-SD+TV applies TV independently to FOD amplitudes evaluated on a discrete spherical grid, the present approach applies TV to SH coefficient maps. Because SH coefficients are orthogonal, this basis may enable more efficient denoising. However, a direct comparison with RUMBA-SD+TV or other methods was beyond the scope of this work. Our objective was to extend CSD/Super-CSD to incorporate spatial information. Results from comparisons across different algorithmic families (e.g., with distinct cost functions or optimization strategies) can be strongly influenced by phantom properties such as the fraction of multi-fiber voxels, the distribution of inter-fiber angles, the proportion of non-dominant fibers, or varying anisotropy levels, as demonstrated in the 'Sparse-wars' study \cite{Canales2019_sparse}. These confounding factors were minimized in the present work by restricting the analysis to a single algorithmic class.

Future work should investigate alternative strategies for constructing spatial priors, including machine learning–based denoisers such as Patch2Self \cite{Patch2Self}. Moreover, the current two-step procedure—first estimating a prior and then re-estimating the FOD—could be generalized into an iterative framework that minimizes a global multi-voxel cost function, jointly penalizing data misfit and spatial discontinuity, potentially through projected gradient descent. We defer this extension to future work, as incorporating the non-negativity constraint into such a formulation requires further study. Another limitation of the present work is that we restricted the analysis to $l_{max}=12$ in the Super-CSD variant, which provided a reasonable trade-off between the angular resolution and the number of parameters. At least this value was required to operate in the super-resolution regime with our datasets, effectively doubling the number of parameters relative to the conventional $l_{max}=8$. Future studies could explore the optimal choice of $l_{max}$ as a function of the number of diffusion gradient directions and the SNR; preliminary unpublished tests (data not shown) suggest that higher values may yield comparable performance, but a systematic evaluation is still needed. 

A limitation of the proposed approach is its partial dependence on the quality of the initial Super-CSD estimates and the TV filtering step. Although the J-invariance calibration mitigates the risk of over-smoothing and the final optimization balances the spatial prior with data fidelity, in cases where the initial CSD estimates are severely degraded by image artifacts, the proposed approach may not fully correct such imperfections. Testing the proposed method under even more constrained protocols (e.g., low b-values and sparse sampling schemes) would be of interest for broader clinical applicability, and we consider this to be a relevant direction for future work. The evaluation of SR$^2$-CSD in large populations with pathological brain tissue remains a key question that should be addressed in future studies.

In our implementation, the only fixed parameter is the regularization hyperparameter $\rho$ that balances the data-fitting and prior terms. In the scaled problem, setting $\rho=1$ ensures comparable contributions from both terms. This choice is justified by considering that our prior is highly informative. It is a spatially denoised (TV-filtered) version of the FOD estimated with Super-CSD. Since the original Super-CSD FOD already minimizes the data-fitting term, its denoised counterpart produces residuals of similar magnitude. Furthermore, by the J-invariance property of the denoiser, these residuals may even better approximate the measurement noise than those from the unregularized solution. However, we acknowledge that this value may not be universally optimal for all voxels, datasets, and SNR levels. In principle, more sophisticated strategies such as cross-validation or L-curve analysis could be employed for adaptive tuning (e.g. \cite{Canales-Rodríguez2021a, Canales-Rodríguez2021b}), but at a considerable computational cost. Our empirical results indicate that fixing $\rho$ at this value already provides robust and stable improvements over reference methods. For completeness, it is worth mentioning that the original CSD method involves a similar regularization hyperparameter $k_1$, which is likewise set to one (see Eq.~(\ref{eq:5a})).

It is important to note that the benefits of SR$^2$-CSD are expected to be most appreciable under realistic noise conditions. In the absence of noise, FOD estimation is intrinsically stable, and the residual noise in the FOD map is negligible. In such cases, the TV-denoised Super-CSD FOD (i.e. the prior) becomes nearly indistinguishable from the undenoised Super-CSD solution, since the denoising strength in our framework depends on the noise level estimated from the FOD map. Consequently, in noise-free data, SR$^2$-CSD would produce results very similar to those of Super-CSD. For this reason, our evaluation focused on SNR levels that are typically encountered in real dMRI acquisitions.

The proposed method will be freely available in the DIPY software library \cite{garyfallidis2014dipy}, thus promoting open access and reproducibility in diffusion MRI research.

\section{Acknowledgments}
This work is supported by the Swiss National Science Foundation (SNSF) under the grant Nbr 205320$\_$204097.

\bibliographystyle{elsarticle-num} 
\bibliography{references.bib}

\clearpage
\appendix
\input{elsarticle/supplementary}

\end{document}

%% file: elsarticle/supplementary.tex
\sisetup{separate-uncertainty=true}
\setcounter{figure}{0}
\renewcommand{\thefigure}{S\arabic{figure}}
\setcounter{table}{0}
\renewcommand{\thetable}{S\arabic{table}}

\title{Supplementary Material}
\date{}
\begin{center}
    \LARGE \text{Supplementary Material}
\end{center}
\vspace{.5cm}

This Supplementary Material provides extended results and additional analyses complementing the main manuscript. Section S1 presents statistical test figures and quantitative metrics from the HARDI phantom experiments at SNR = 10 and 15, evaluated on both MPPCA-denoised and raw data using single-shell and multi-shell acquisitions. Section S2 reports detailed reproducibility analyses for the test-retest dataset, including both the full acquisition with 96 diffusion directions and a subsampled dataset with 45 directions.

\section*{S1. HARDI Phantom Results: Statistical Tests and Quantitative Metrics}

\begin{figure*}[h!]
  \centering
  \includegraphics[width=.9\textwidth]{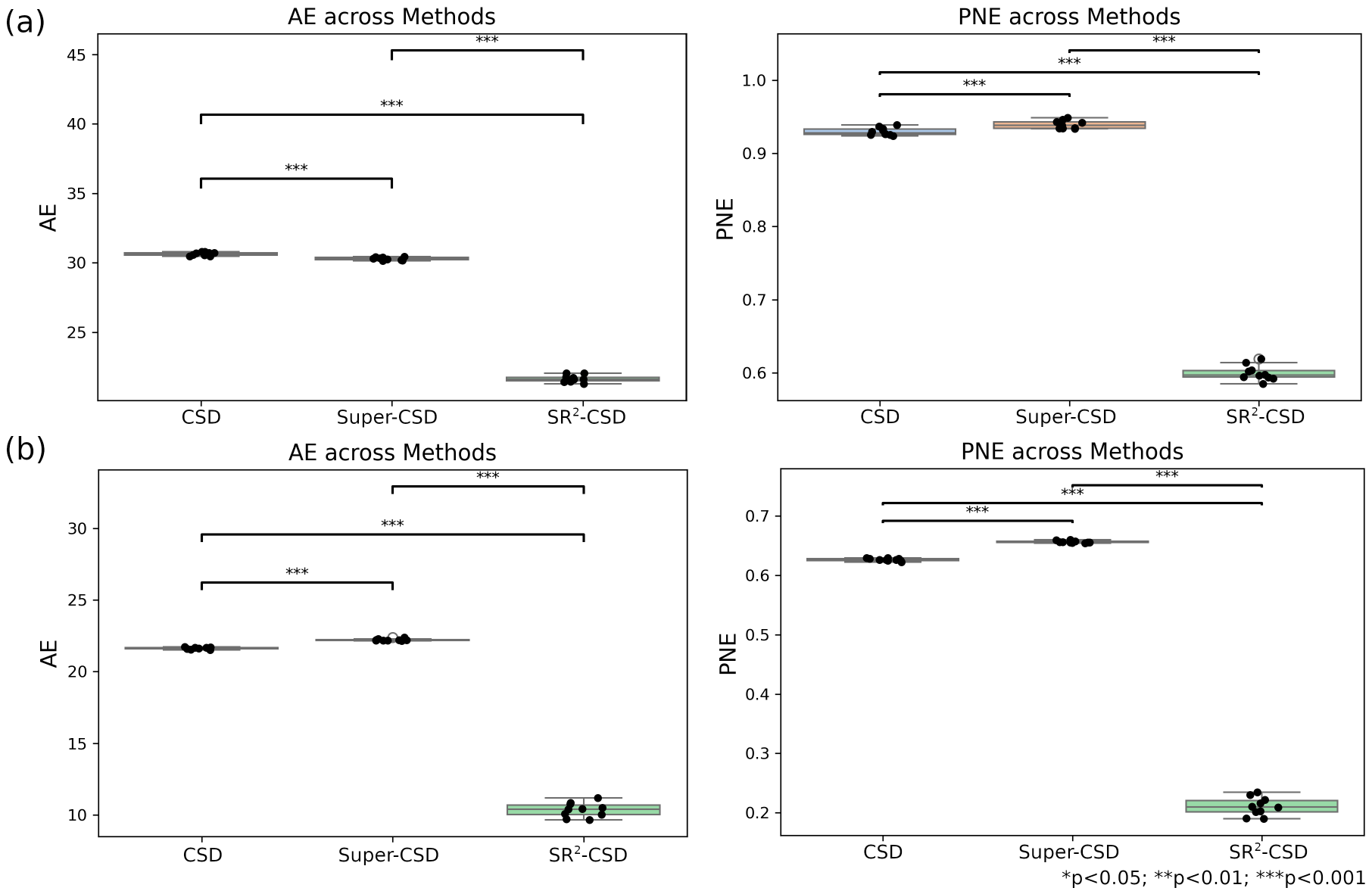}
  \caption{Reconstruction accuracy on the HARDI phantom (SNR = 10, 10 noise realizations, for single-shell single-tissue reconstruction). (a) Without MPPCA denoising. (b) After MPPCA denoising. Boxplots show Angular Error (AE) and Peak Number Error (PNE) for CSD ($l_{max}=8$), Super-CSD ($l_{max}=12$), and SR$^2$-CSD ($l_{max}=12$). Statistical significance assessed via paired t-tests with Bonferroni correction. Asterisks indicate significance levels: *$p < 0.05$; **$p < 0.01$; ***$p < 0.001$.}
\end{figure*}

\begin{figure*}[h!]
  \centering
  \includegraphics[width=.9\textwidth]{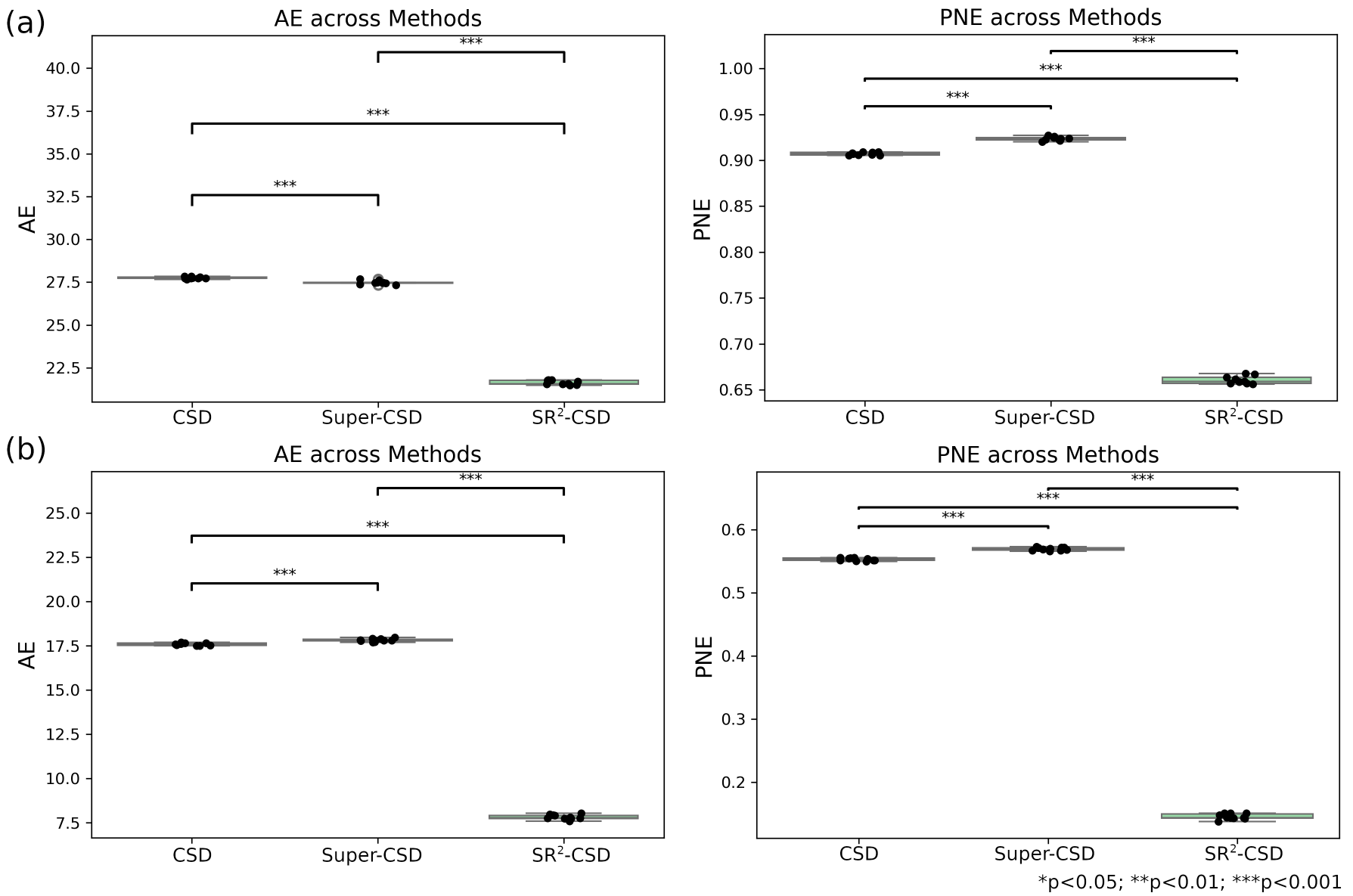}
  \caption{Reconstruction accuracy on the HARDI phantom (SNR = 15, 10 noise realizations, for single-shell single-tissue reconstruction). (a) Without MPPCA denoising. (b) After MPPCA denoising. Boxplots show Angular Error (AE) and Peak Number Error (PNE) for CSD ($l_{max}=8$), Super-CSD ($l_{max}=12$), and SR$^2$-CSD ($l_{max}=12$). Statistical significance assessed via paired t-tests with Bonferroni correction. Asterisks indicate significance levels: *$p < 0.05$; **$p < 0.01$; ***$p < 0.001$.}
\end{figure*}

\begin{figure}[h!]
  \centering
  \includegraphics[width=.9\textwidth]{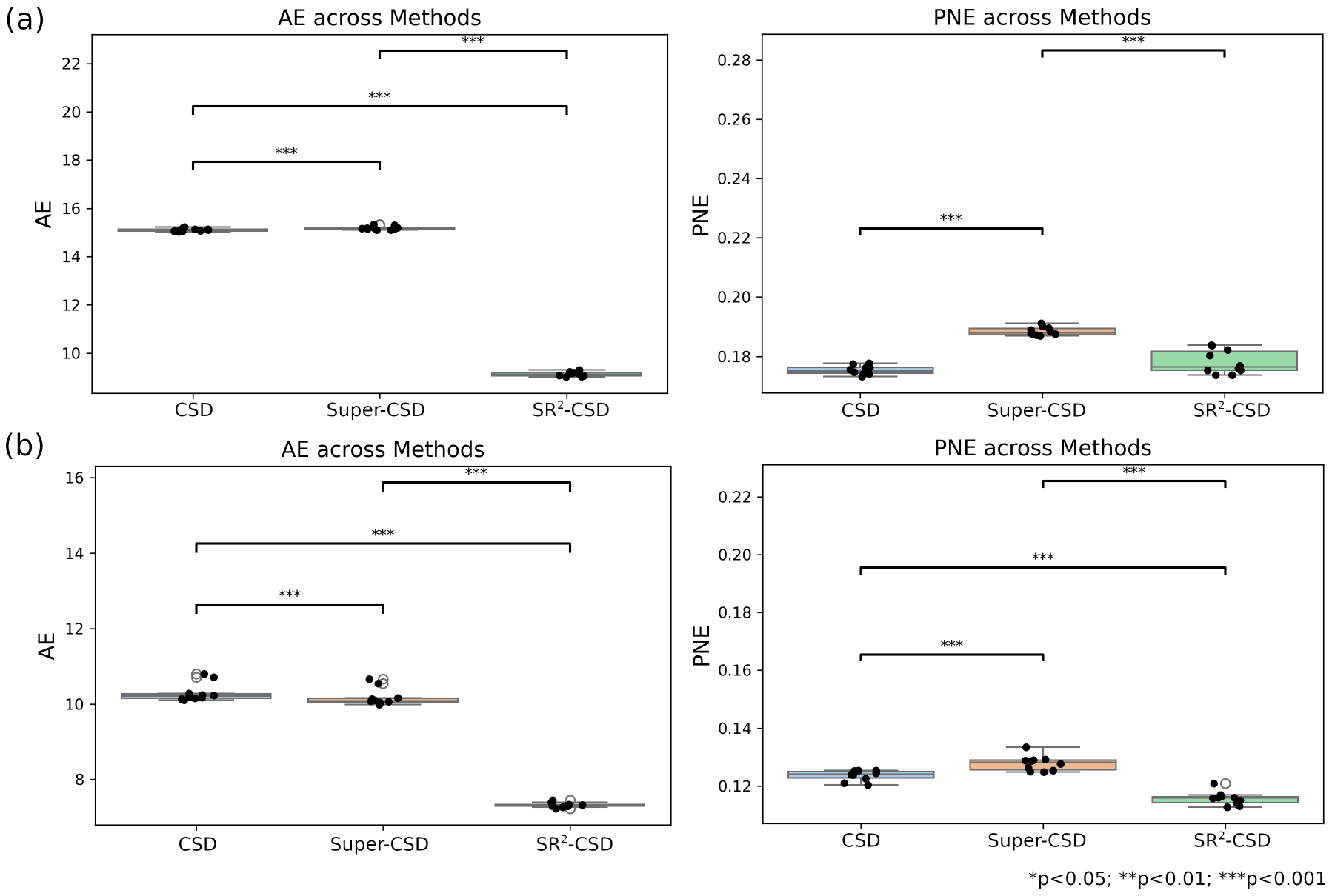}
  \caption{Reconstruction accuracy on the HARDI phantom (SNR = 10, 10 noise realizations, for multi-shell multi-tissue reconstruction). (a) Without MPPCA denoising. (b) After MPPCA denoising. Boxplots show Angular Error (AE) and Peak Number Error (PNE) for CSD ($l_{max}=8$), Super-CSD ($l_{max}=12$), and SR$^2$-CSD ($l_{max}=12$). Statistical significance assessed via paired t-tests with Bonferroni correction. Asterisks indicate significance levels: *$p < 0.05$; **$p < 0.01$; ***$p < 0.001$.}
\end{figure}

\begin{figure}[h!]
  \centering
  \includegraphics[width=.9\textwidth]{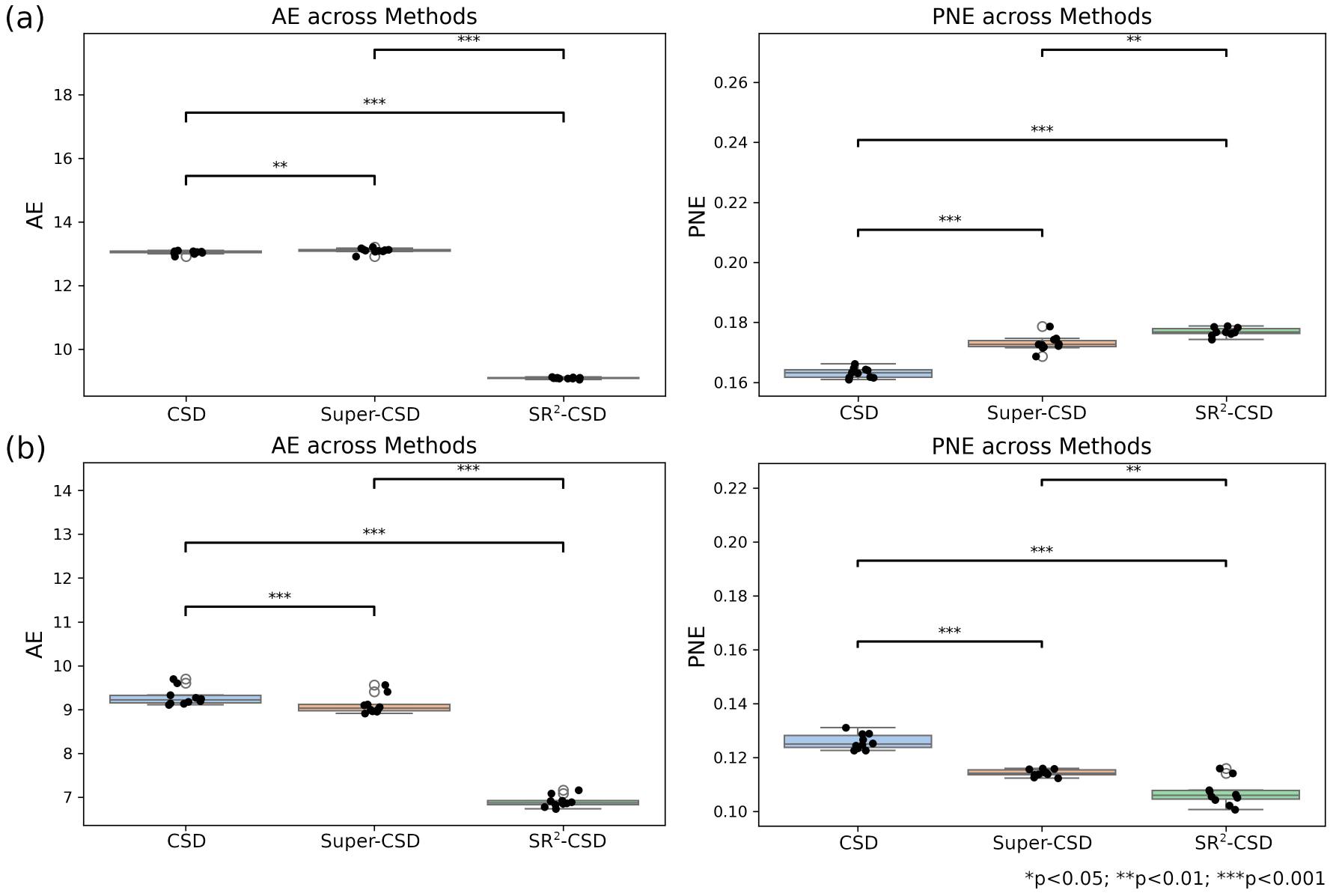}
 \caption{Reconstruction accuracy on the HARDI phantom (SNR = 15, 10 noise realizations, for multi-shell multi-tissue reconstruction). (a) Without MPPCA denoising. (b) After MPPCA denoising. Boxplots show Angular Error (AE) and Peak Number Error (PNE) for CSD ($l_{max}=8$), Super-CSD ($l_{max}=12$), and SR$^2$-CSD ($l_{max}=12$). Statistical significance assessed via paired t-tests with Bonferroni correction. Asterisks indicate significance levels: *$p < 0.05$; **$p < 0.01$; ***$p < 0.001$.}
\end{figure}

\begin{table}[h]
\centering
\caption{HARDI phantom with non-denoised data for the single-shell single-tissue reconstruction. Angular Error (AE) and Peak Number Error (PNE) metrics were computed for CSD with $l_{max} = 8$, and Super-CSD and SR$^2$-CSD with $l_{max} = 12$. The mean and standard deviation were calculated from estimates based on 10 different noise realizations. The reported means correspond to the average of per-realization mean values, and the standard deviations reflect the variability across these repetitions. The optimal results for each SNR are highlighted in bold.}
\label{hardi_tables_ssst_raw}
\vspace{0.2cm}
\begin{tabular}{l@{\hspace{1.5em}}l@{\hspace{2em}}c@{\hspace{2em}}c}

\toprule
\textbf{SNR} & \textbf{Method} & \textbf{AE (degree)} & \textbf{PNE} \\
\midrule
\multirow{3}{*}{10}

 & CSD & 30.646 $\pm$ 0.115 & 0.930 $\pm$ 0.005  \\ 
 & Super-CSD & 30.299 $\pm$ 0.096 & 0.939 $\pm$ 0.005 \\ 
 & SR$^2$-CSD & \textbf{21.672 $\pm$ 0.239} & \textbf{0.599 $\pm$ 0.009} \\
\midrule 
\multirow{3}{*}{15}
 & CSD & 27.758 $\pm$ 0.054 & 0.907 $\pm$ 0.002  \\ 
 & Super-CSD & 27.482 $\pm$ 0.102 & 0.924 $\pm$ 0.002 \\ 
 & SR$^2$-CSD & \textbf{21.632 $\pm$ 0.118} & \textbf{0.660 $\pm$ 0.004} \\
\midrule
\multirow{3}{*}{30}
 & CSD & 24.538 $\pm$ 0.083 & 0.855 $\pm$ 0.002 \\ 
 & Super-CSD & 25.142 $\pm$ 0.075 & 0.895 $\pm$ 0.003 \\ 
 & SR$^2$-CSD & \textbf{16.764 $\pm$ 0.073} & \textbf{0.501 $\pm$ 0.002} \\
\bottomrule
\end{tabular}
\end{table}

\begin{table}[h]
\centering
\caption{HARDI phantom with denoised data for the single-shell single-tissue reconstruction. Angular Error (AE) and Peak Number Error (PNE) metrics were computed for CSD with $l_{max} = 8$, and Super-CSD and SR$^2$-CSD with $l_{max} = 12$. The mean and standard deviation were calculated from estimates based on 10 different noise realizations. The reported means correspond to the average of per-realization mean values, and the standard deviations reflect the variability across these repetitions. The optimal results for each SNR are highlighted in bold.}
\label{hardi_tables_ssst_denoised}
\vspace{0.2cm}
\begin{tabular}{l@{\hspace{1.5em}}l@{\hspace{2em}}c@{\hspace{2em}}c}
\toprule
\textbf{SNR} & \textbf{Method} & \textbf{AE (degree)} & \textbf{PNE} \\
\midrule
\multirow{3}{*}{10}
 & CSD & 21.635 $\pm$ 0.068 & 0.626 $\pm$ 0.002 \\ 
 & Super-CSD & 22.225 $\pm$ 0.072 & 0.656 $\pm$ 0.002 \\ 
 & SR$^2$-CSD & \textbf{10.363 $\pm$ 0.497} & \textbf{0.210 $\pm$ 0.015} \\
\midrule 
\multirow{3}{*}{15}
 & CSD & 17.592 $\pm$ 0.073 & 0.553 $\pm$ 0.002 \\ 
 & Super-CSD & 17.822 $\pm$ 0.084 & 0.570 $\pm$ 0.002 \\ 
 & SR$^2$-CSD & \textbf{7.831 $\pm$ 0.133} & \textbf{0.145 $\pm$ 0.004} \\
\midrule
\multirow{3}{*}{30}
 & CSD & 14.531 $\pm$ 0.097 & 0.468 $\pm$ 0.004 \\ 
 & Super-CSD & 14.877 $\pm$ 0.068 & 0.484 $\pm$ 0.003 \\ 
 & SR$^2$-CSD & \textbf{7.624 $\pm$ 0.057} & \textbf{0.140 $\pm$ 0.001} \\
\bottomrule
\end{tabular}
\end{table}

\begin{table}[h]
\centering
\caption{HARDI phantom with non-denoised data for the multi-shell multi-tissue reconstruction. Angular Error (AE) and Peak Number Error (PNE) metrics were computed for CSD with $l_{max} = 8$, and Super-CSD and SR$^2$-CSD with $l_{max} = 12$. The mean and standard deviation were calculated from estimates based on 10 different noise realizations. The reported means correspond to the average of per-realization mean values, and the standard deviations reflect the variability across these repetitions. The optimal results for each SNR are highlighted in bold.}
\label{hardi_tables_msmt_raw}
\vspace{0.2cm}
\begin{tabular}{l@{\hspace{1.5em}}l@{\hspace{2em}}c@{\hspace{2em}}c}
\toprule
\textbf{SNR} & \textbf{Method} & \textbf{AE (degree)} & \textbf{PNE} \\
\midrule
\multirow{3}{*}{10}
 & CSD & 15.106 $\pm$ 0.065 & \textbf{0.176 $\pm$ 0.002} \\ 
 & Super-CSD & 15.183 $\pm$ 0.074 & 0.189 $\pm$ 0.002 \\ 
 & SR$^2$-CSD & \textbf{9.130 $\pm$ 0.091} &  0.178 $\pm$ 0.003 \\
\midrule
\multirow{3}{*}{15}
 & CSD & 13.041 $\pm$ 0.051 & \textbf{0.163 $\pm$ 0.002} \\ 
 & Super-CSD & 13.099 $\pm$ 0.075 & 0.173 $\pm$ 0.003 \\ 
 & SR$^2$-CSD & \textbf{9.089 $\pm$ 0.021} & 0.176 $\pm$ 0.001 \\
\midrule
\multirow{3}{*}{30}
 & CSD & 10.246 $\pm$ 0.045 & 0.143 $\pm$ 0.002 \\ 
 & Super-CSD & 10.158 $\pm$ 0.053 & 0.131 $\pm$ 0.001 \\ 
 & SR$^2$-CSD & \textbf{7.554 $\pm$ 0.044} &  \textbf{0.126 $\pm$ 0.001}\\
\bottomrule
\end{tabular}
\end{table}

\begin{table}[h]
\centering
\caption{HARDI phantom with denoised data for the multi-shell multi-tissue reconstruction. Angular Error (AE) and Peak Number Error (PNE) metrics were computed for CSD with $l_{max} = 8$, and Super-CSD and SR$^2$-CSD with $l_{max} = 12$. The mean and standard deviation were calculated from estimates based on 10 different noise realizations. The reported means correspond to the average of per-realization mean values, and the standard deviations reflect the variability across these repetitions. The optimal results for each SNR are highlighted in bold.}
\label{hardi_tables_msmt_denoised}
\vspace{0.2cm}
\begin{tabular}{l@{\hspace{1.5em}}l@{\hspace{2em}}c@{\hspace{2em}}c}
\toprule
\textbf{SNR} & \textbf{Method} & \textbf{AE (degree)} & \textbf{PNE} \\
\midrule
\multirow{3}{*}{10}
 & CSD & 10.303 $\pm$ 0.232 & 0.124 $\pm$ 0.002 \\ 
 & Super-CSD & 10.181 $\pm$ 0.220 & 0.128 $\pm$ 0.002 \\ 
 & SR$^2$-CSD & \textbf{7.327 $\pm$ 0.065} & \textbf{0.115 $\pm$ 0.002} \\
\midrule
\multirow{3}{*}{15}
 & CSD & 9.291 $\pm$ 0.193 & 0.126 $\pm$ 0.003 \\ 
 & Super-CSD & 9.107 $\pm$ 0.201 & 0.114 $\pm$ 0.001 \\ 
 & SR$^2$-CSD & \textbf{6.901 $\pm$ 0.124} & \textbf{0.106 $\pm$ 0.004} \\
\midrule
\multirow{3}{*}{30} 
 & CSD & 9.610 $\pm$ 0.080 & 0.190 $\pm$ 0.004 \\ 
 & Super-CSD & 9.636 $\pm$ 0.105 & 0.136 $\pm$ 0.002 \\ 
 & SR$^2$-CSD & \textbf{6.488 $\pm$ 0.102} & \textbf{0.095 $\pm$ 0.004} \\
\bottomrule
\end{tabular}
\end{table}

\clearpage
\section*{S2. Test–Retest Reproducibility Analysis}
\subsection*{Subsampling from 96 to 45 DW Directions}

\begin{figure}[h!]
  \centering
  \includegraphics[width=0.8\textwidth]{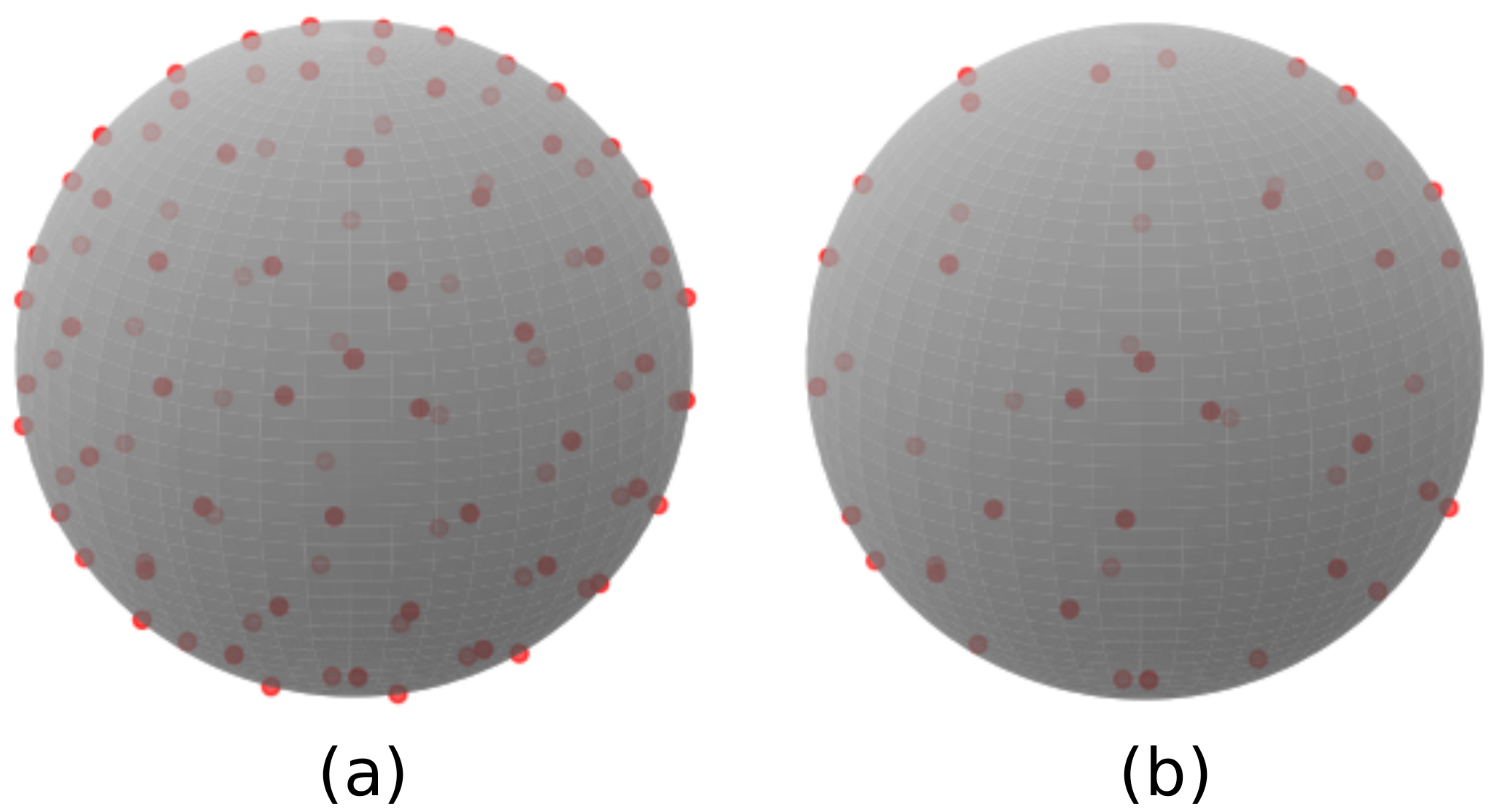}
  \caption{Diffusion gradient sampling schemes: (a) Full set of 96 gradient directions; (b) subsampled 45 gradient directions.}
\label{sampling_sphere}
\end{figure}

Originally, the test-retest dMRI data was acquired using 96 unique gradient directions (excluding 10 of $b = 0$ volumes). To evaluate reconstruction methods under clinically feasible conditions, we subsampled the dataset to obtain 45 gradient directions. The two sampling schemes are shown in Figure \ref{sampling_sphere}.

\begin{figure}[H]
  \centering
  \includegraphics[width=1\textwidth]{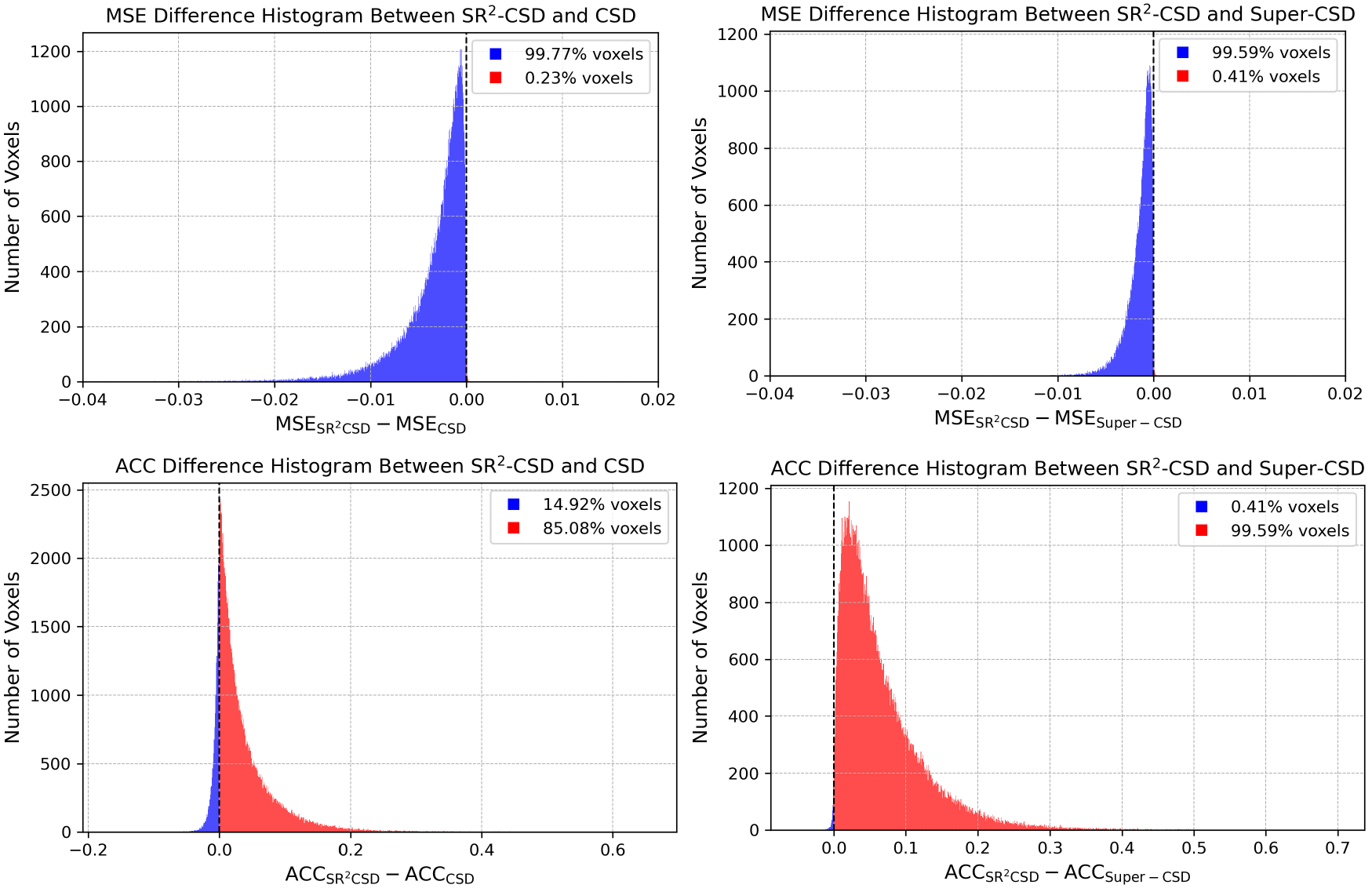}
  \caption{Voxel-wise difference histograms comparing SR$^{2}$-CSD with CSD ($l_{max}=8$) and Super-CSD ($l_{max}=12$) for Subject 3 using MPPCA-denoised data acquired with the full scheme of 96 gradient directions. For more details see Figure 7 of the main manuscript. Top row: MSE differences, where negative values indicate lower MSE for SR$^{2}$-CSD. Bottom row: ACC differences, where positive values indicate higher ACC for SR$^{2}$-CSD. Legends report the proportion of voxels favoring each method. Across both comparisons, SR$^{2}$-CSD shows lower MSE and higher ACC in most voxels.}
\label{metric_difference_histograms}
\end{figure}

To further examine the voxel-wise differences between methods for the results reported in Figure 7 of the main manuscript, we generated histograms of the mean squared error (MSE) and angular correlation coefficient (ACC) difference maps between SR$^2$-CSD and the two reference methods CSD and Super-CSD for the same subject (see Figure \ref{metric_difference_histograms}). Voxel-wise differences of MSE and ACC were computed within the white matter mask. Negative values in MSE histograms (top row) indicate voxels where SR$^2$-CSD yields lower MSE than CSD and Super-CSD. In contrast, positive values in the ACC histograms (bottom row) indicate voxels where SR$^2$-CSD achieves higher ACC than the reference. The results confirm that SR$^2$-CSD achieves lower MSE and higher ACC than both CSD and Super-CSD in the vast majority of voxels, supporting its improved reproducibility.

More detailed results from the six subjects studied are reported in Tables \ref{subjects_table_45_dir} and \ref{subjects_table_96_dir}. Table \ref{subjects_table_45_dir} summarizes the reproducibility metrics with the subsampled 45-direction dataset, while Table \ref{subjects_table_96_dir} reports the results using the full 96-direction dataset. For all subjects, SR$^{2}$-CSD consistently achieved a lower MSE and a higher angular correlation coefficient (ACC) compared to CSD and Super-CSD. Moreover, SR$^{2}$-CSD produced the lowest angular error (AE) in four and all subjects for the subsampled and full datasets, respectively. Similarly, SR$^{2}$-CSD obtained the lowest peak number error (PNE) values in four and five subjects for the subsampled and full datasets.

\setlength{\tabcolsep}{8pt}
\begin{table}[H]
\centering
\caption{Results from the reproducibility study using \textbf{the subsampled dataset with 45 DW directions} for CSD with $l_{max}=8$, and Super-CSD and SR$^2$-CSD with $l_{max}=12$. Mean squared error (MSE), angular correlation coefficient (ACC), angular error (AE), and peak number error (PNE) metrics were calculated using the two Fiber Orientation Distribution (FOD) estimates from the two scans per subject. The optimal metrics for each subject are highlighted in bold.}
\vspace{.3em}
\label{subjects_table_45_dir}
\begin{tabular}{c l c c c c}
\toprule
\textbf{Subject} & \textbf{Method} & \textbf{MSE} & \textbf{ACC} & \textbf{AE} & \textbf{PNE} \\
\midrule
\multirow{3}{*}{1}
    & CSD & 0.322 & 0.838 & \textbf{21.29} & 0.212 \\
    & Super-CSD & 0.650 & 0.675 & 25.65 & 0.247 \\
    & SR$^2$-CSD & \textbf{0.183} & \textbf{0.908} & 24.38 & \textbf{0.207} \\
\midrule 
\multirow{3}{*}{2}
    & CSD & 0.330 & 0.834 & 21.79 & 0.220 \\
    & Super-CSD & 0.707 & 0.646 & 26.83 & \textbf{0.191} \\
    & SR$^2$-CSD & \textbf{0.104} & \textbf{0.948} & \textbf{20.95} & 0.223 \\
\midrule 
\multirow{3}{*}{3}
    & CSD & 0.376 & 0.812 & 24.46 & 0.228 \\
    & Super-CSD & 0.835 & 0.583 & 26.98 & 0.308 \\
    & SR$^2$-CSD & \textbf{0.255} & \textbf{0.872} & \textbf{20.27} & \textbf{0.221} \\
\midrule 
\multirow{3}{*}{4}
    & CSD & 0.333 & 0.837 & 21.54 & 0.229 \\
    & Super-CSD & 0.591 & 0.706 & 24.52 & \textbf{0.222} \\
    & SR$^2$-CSD & \textbf{0.267} & \textbf{0.871} & \textbf{21.49} & 0.229 \\
\midrule 
\multirow{3}{*}{5}
    & CSD & 0.299 & 0.850 & 21.63 & 0.225 \\
    & Super-CSD & 0.596 & 0.702 & 24.65 & 0.211 \\
    & SR$^2$-CSD & \textbf{0.230} & \textbf{0.885} & \textbf{21.58} & \textbf{0.207} \\
\midrule 
\multirow{3}{*}{6}
    & CSD & 0.399 & 0.800 & \textbf{22.94} & 0.218 \\
    & Super-CSD & 0.814 & 0.593 & 27.54 & 0.243 \\
    & SR$^2$-CSD & \textbf{0.211} & \textbf{0.895} & 25.77 & \textbf{0.197} \\
\bottomrule
\end{tabular}
\end{table}

\setlength{\tabcolsep}{8pt}
\begin{table}[H]
\centering
\caption{Results from the reproducibility study using \textbf{the full dataset with 96 DW directions} for CSD with $l_{max}=8$, and Super-CSD and SR$^2$-CSD with $l_{max}=12$. Mean squared error (MSE), angular correlation coefficient (ACC), angular error (AE), and peak number error (PNE) metrics were calculated using the two Fiber Orientation Distribution (FOD) estimates from the two scans per subject. The optimal metrics for each subject are highlighted in bold.}
\label{subjects_table_96_dir}
\vspace{.3em}
\begin{tabular}{c l c c c c}
\toprule
\textbf{Subject} & \textbf{Method} & \textbf{MSE} & \textbf{ACC} & \textbf{AE} & \textbf{PNE} \\
\midrule
\multirow{3}{*}{1}
    & CSD & 0.265 & 0.867 & 19.30 & 0.188 \\
    & Super-CSD & 0.342 & 0.828 & 20.68 & 0.203 \\
    & SR$^2$-CSD & \textbf{0.180} & \textbf{0.910} & \textbf{17.53} & \textbf{0.184} \\
\midrule
\multirow{3}{*}{2}
    & CSD & 0.277 & 0.861 & 20.07 & 0.190 \\
    & Super-CSD & 0.362 & 0.819 & 21.36 & 0.197 \\
    & SR$^2$-CSD & \textbf{0.179} & \textbf{0.912} & \textbf{18.09} & \textbf{0.188} \\
\midrule 
\multirow{3}{*}{3}
    & CSD & 0.295 & 0.852 & 19.65 & 0.182 \\
    & Super-CSD & 0.364 & 0.817 & 20.69 & 0.202 \\
    & SR$^2$-CSD & \textbf{0.193} & \textbf{0.903} & \textbf{17.33} & \textbf{0.175} \\
\midrule
\multirow{3}{*}{4}
    & CSD & 0.266 & 0.871 & 19.01 & 0.191 \\
    & Super-CSD & 0.337 & 0.835 & 20.10 & 0.208 \\
    & SR$^2$-CSD & \textbf{0.173} & \textbf{0.918} & \textbf{16.70} & \textbf{0.184} \\
\midrule
\multirow{3}{*}{5}
    & CSD & 0.234 & 0.883 & 19.75 & 0.209 \\
    & Super-CSD & 0.316 & 0.842 & 20.99 & 0.217 \\
    & SR$^2$-CSD & \textbf{0.179} & \textbf{0.910} & \textbf{17.86} & \textbf{0.206} \\
\midrule 
\multirow{3}{*}{6}
    & CSD & 0.325 & 0.837 & 20.74 & \textbf{0.194} \\
    & Super-CSD & 0.407 & 0.796 & 21.74 & 0.205 \\
    & SR$^2$-CSD & \textbf{0.224} & \textbf{0.888} & \textbf{18.83} & 0.203 \\
\bottomrule
\end{tabular}
\end{table}

\section*{S3. Diffusion-Simulated Connectivity (DiSCo) Dataset Tractography Results}

\begin{table}[h!]
\setlength{\tabcolsep}{12pt}
\centering
\caption{Mean and standard deviation of the Pearson correlation coefficient ($r$), calculated between the estimated and ground-truth connectivity matrices, for the CSD, Super-CSD, and SR$^2$-CSD methods across multiple SNR levels using 10 independent noise realizations per method. The optimal results for each SNR are highlighted in bold.}
\vspace{1em}
\label{tab:supp_corr_stats}
\renewcommand{\arraystretch}{1.3}
\begin{tabular}{c|ccc}
\hline
\textbf{SNR} & \textbf{CSD} & \textbf{Super-CSD} & \textbf{SR$^2$-CSD} \\
\hline
\rowcolor{gray!10} 10 & 0.8788 $\pm$ 0.0097 & 0.8814 $\pm$ 0.0147 & \textbf{0.8919 $\pm$ 0.0109} \\
20 & 0.9063 $\pm$ 0.0057 & 0.9039 $\pm$ 0.0084 & \textbf{0.9202 $\pm$ 0.0053} \\
\rowcolor{gray!10} 30 & 0.9161 $\pm$ 0.0034 & 0.9095 $\pm$ 0.0039 & \textbf{0.9227 $\pm$ 0.0029} \\
50 & 0.9188 $\pm$ 0.0039 & 0.9144 $\pm$ 0.0050 & \textbf{0.9230 $\pm$ 0.0032} \\
\hline
\end{tabular}
\end{table}
